\documentclass[twocolumn]{aastex62}
\graphicspath{{./}{figures/}}

\received{}
\revised{}
\accepted{}
\submitjournal{ApJ}

\shorttitle{Origin of hard X-ray emissions during the 2014 September 1 event}
\shortauthors{Wu et al.}

\begin{document}

\title{On the origin of hard X-ray emissions from the behind-the-limb flare on 2014 September 1}

\correspondingauthor{Yihong Wu}
\email{yhwyihongwu@gmail.com}

\author[0000-0003-1660-0513]{Yihong Wu}
\affiliation{Institut de Recherche en Astrophysique et Planétologie, Université de Toulouse III (UPS), France}
\affiliation{Centre National de la Recherche Scientifique, UMR 5277, Toulouse, France}
\affiliation{Leibniz-Institut f\"{u}r Astrophysik Potsdam (AIP), Potsdam, Germany}

\author[0000-0003-4039-5767]{Alexis P. Rouillard}
\affiliation{Institut de Recherche en Astrophysique et Planétologie, Université de Toulouse III (UPS), France}
\affiliation{Centre National de la Recherche Scientifique, UMR 5277, Toulouse, France}

\author{Athanasios Kouloumvakos}
\affiliation{Institut de Recherche en Astrophysique et Planétologie, Université de Toulouse III (UPS), France}
\affiliation{Centre National de la Recherche Scientifique, UMR 5277, Toulouse, France}

\author[0000-0002-3298-2067]{Rami Vainio}
\affiliation{University of Turku, Turku, Finland}

\author[0000-0001-9325-6758]{Alexandr N. Afanasiev}
\affiliation{University of Turku, Turku, Finland}

\author[0000-0002-0074-4048]{Illya Plotnikov}
\affiliation{Institut de Recherche en Astrophysique et Planétologie, Université de Toulouse III (UPS), France}

\author{Ronald J. Murphy}
\affiliation{Space Science Division, Naval Research Laboratory, Washington DC 20375, USA}

\author{Gottfried J. Mann}
\affiliation{Leibniz-Institut f\"{u}r Astrophysik Potsdam (AIP), Potsdam, Germany}

\author[0000-0003-1439-3610]{Alexander Warmuth}
\affiliation{Leibniz-Institut f\"{u}r Astrophysik Potsdam (AIP), Potsdam, Germany}



\begin{abstract}

The origin of hard X-rays and $\gamma$-rays emitted from the solar atmosphere during occulted solar flares is still debated. The hard X-ray emissions could come from flaring loop tops rising above the limb or Coronal Mass Ejections (CME) shock waves, two by-products of energetic solar storms. For the shock scenario to work, accelerated particles must be released on magnetic field lines rooted on the visible disk and precipitate. We present a new Monte Carlo code that computes particle acceleration at shocks propagating along large coronal magnetic loops. A first implementation of the model is carried out for the 2014 September 1 event and the modeled electron spectra are compared with those inferred from Fermi Gamma-ray Burst Monitor (GBM) measurements. When particle diffusion processes are invoked our model can reproduce the hard electron spectra measured by GBM nearly ten minutes after the estimated on-disk hard X-rays appear to have ceased from the flare site.\\

\end{abstract}



\section{Introduction}
\label{sec:intro}
   \subsection{The 2014 September 1 event}
   \label{sec:event}
  
NASA's Fermi-Large Area Telescope (LAT) \citep{2009ApJ...697.1071A} detected more than 30 solar eruptive events with late phase $>$ 100 MeV $\gamma$-ray emission during its ten-year mission, among which the 2014 September 1 event was one of largest. The $\gamma$-ray emission was generated from decay of pions, which were likely created by interactions of high-energy protons that impacted chromospheric material on the visible disk. The NaI and BGO detectors on Fermi's second instrument, the Gamma-ray Burst Monitor (GBM) \citep{2009ApJ...702..791M}, also observed hard X-ray emission in the MeV range during this event, which originated from thin or thick-target electron bremsstrahlung \citep{2018ApJ...869..182S}.\\

Several studies have focused on this solar event launched 36$^\circ$ beyond the East Limb of the Sun \citep[e.g.,][]{2017ApJ...835..219A,2017A&A...608A..43P, 2018SoPh..293..133G,2018ApJ...867..122J,2018ApJ...865...99P,2018ApJ...869..182S,2020SoPh..295...18G}. While the source of $>$ 100 MeV $\gamma$-ray emission is observed on the visible disk, the origin of the hard X-ray emission is not clear. Two strong candidates are the flare loop tops via electron thin-target emission and thick-target emission resulting from electrons impacting the visible disk or perhaps a combination of both. Since the magnetic footpoints of flaring loops associated with the CME eruption were occulted, it was argued by \citet{2017A&A...608A..43P} that flare-accelerated electrons were unlikely to be responsible for the measured hard X-ray emission. By means of detailed 3-D coronal modeling \citet{2017A&A...608A..43P} showed that the onset of the hard-X ray and $\gamma$-ray emissions occurred when the CME-driven coronal shock became magnetically connected to the visible disk. This suggested that electrons and protons accelerated at the coronal shock had a means to propagate toward the visible disk and impact the chromosphere to produce high-energy radiation. For this event, the shock is a strong candidate for particle acceleration. The moving shock wave front could be an important electron accelerator far from the flare site, as illustrated in the studies of \citet{1999ApJ...519..864K} and \citet{2012ApJ...752...44}. \citet{1999ApJ...519..864K} has presented a good correlation between electron beams detected by Wind and extreme ultraviolet (EUV) waves observed by the Solar and Heliospheric Observatory (SoHO) spacecraft. \citet{2012ApJ...752...44} argued that 0.67-3.08 MeV electrons measured near Earth during the 2011 March 21 event could have been accelerated by the CME shock. The flaring site was on the far side of the Sun as viewed from Earth (W132$^\circ$).\\

\citet{2017ApJ...835..219A} and \citet{2018SoPh..293..133G} showed that the time history of the hard X-rays measured by Reuven Ramaty High Energy Solar Spectroscopic Imager (RHESSI), Fermi/GBM and Konus-Wind matched closely that of the microwaves measured by the Radio Solar Telescope Network. Since microwaves are likely produced via Synchrotron emission of relativistic electrons in the flaring loops, it was argued that the hard X-rays must also be produced in this region. \citet{2018ApJ...865...99P} presented a detailed physics-based modeling of the event concluding that thin-target processes occurring at the loop tops could explain the hard X-ray spectra.\\

\begin{figure}[hbtp!]
    \epsscale{1.2}
    \plotone{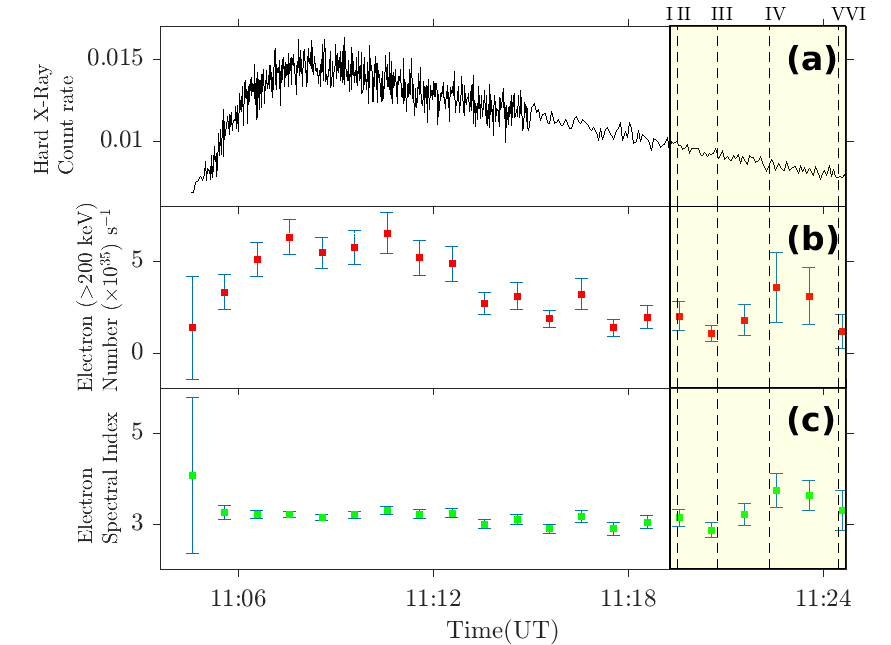}
    \caption{Time series of the arbitrarily scaled 100-300 keV hard X-ray count rate observed by Fermi GBM (Panel a), the average number of $>$ 200 keV electrons (producing the hard X-ray emission) per second in each 1 minute interval with $\pm1\sigma$ statistical uncertainties (Panel b), and $>$ 200 keV electron (producing the hard X-ray emission) power-law spectral indices with $\pm1\sigma$ statistical uncertainties (Panel c). Times when the shock was intersecting FL 2305 are shown as yellow shaded areas. The shock reaches locations I-VI at 11:19:19, 11:19:31, 11:20:45, 11:22:21, 11:24:29 and 11:24:42 UT respectively. The times that the shock crosses locations I and VI on FL 2305 are indicated by vertical solid lines. The times that the shock crosses locations II, III, IV and V on FL 2305 are indicated by vertical dashed lines.
    \label{fig:number_index}}
\end{figure}

In contrast, \citet{2017ApJ...835..219A} found that the electrons producing the $>$ 1 GHz emission had a power-law index of about 3.1, this is also similar to the value found by \citet{2018ApJ...869..182S} for electrons producing the bremsstrahlung in thick target bremsstrahlung. They suggest, at face value, that the electrons emitting at microwave come from the same accelerator as those emitting via thick-target bremsstrahlung. \citet{2018ApJ...869..182S} also argued that, due to the absence of rotational modulation in RHESSI detector 9, $>$ 20 keV hard X-ray emission must have been distributed over a much broader region than the RHESSI source imaged in the 6-12 keV range shown in \citet{2017ApJ...835..219A}. In addition, RHESSI provided only a limited coverage since it was in the South Atlantic Anomaly (SAA) from 10:55:00 to 11:11:00 UT. Furthermore, the NaI detectors on GBM measured enhanced 10-25 keV emission about 2 minutes after the hard X-ray onset observed by the Solar Assembly for X-rays (SAX) on MESSENGER \citep{2007SSRv..131..393S}. \citet{2018ApJ...869..182S} suggested this might be due to the emergence of the flare's coronal hard X-ray source $2\times10^5$ km above the East Limb. Higher-energy emission detected into the MeV range started as the 10-25 keV X-ray flux from the flare decreased. It is possible that this higher-energy emission came from the visible disk. We assume that the hard X-ray source extends over tens of heliographic degrees on the visible disk, far beyond the LAT centroid estimated for this event.\\ 

We conclude that the origin of the hard X-rays is unclear and that further studies must quantify the acceleration and propagation of energetic electrons in both the flaring regions and at the shock waves driven by the emerging CME. In the present study we search for mechanisms that could sustain electron acceleration and hard X-ray production over tens of heliographic degrees on the visible disk via thick target bremsstrahlung in the chromosphere. For that to happen, magnetic connection must be established between the visible disk and the candidate accelerator. We will model electron acceleration to high energies during the lateral motion of the shock as it propagated in the corona toward the visible disk. The aims of this paper are twofold: firstly to present a new modeling framework to address particle acceleration at a shock propagating through coronal loops, and secondly to combine this new model with a shock wave reconstruction carried out for a CME event that erupted on the far side of the Sun and during which a delayed and extended hard X-ray emission was measured by a near-Earth spacecraft.\\

Several studies have looked at the link between the timing of the hard X-ray emission, the propagation of the EUV waves and the onset of the type II burst emission \citep[e.g.,][]{1999A&A...343..287K, 2006A&A...448..739V}. The formation of a quasi-perpendicular shock detected as the type II burst onset is sometimes accompanied by hard X-ray enhancements, e.g. the 1993 September 27 and the 1994 July 7 events \citep{1999A&A...343..287K} as well as the 2003 November 3 event \citep{2006A&A...448..739V}. For these flares visible from Earth, the strong hard X-ray emissions are likely dominated by the bright sources at the chromospheric footpoints and the flaring loop tops. Any contributions to the hard X-ray emission from electrons accelerated by a concomitant CME shock would be extremely hard to be isolated from the strong flare component. Hard X-ray flares with occulted chromospheric sources (i.e. not visible from Earth), such as the event studied in the present paper, are better suited to investigate other electron acceleration mechanisms that may contribute to hard X-ray emissions particularly during the late phase of a flare-CME event.\\

Figure \ref{fig:number_index}~a provides the time history of the arbitrarily scaled 100-300 keV hard X-ray count rate measured by Fermi GBM. The spectra from Fermi GBM were fitted using a thick target electron-proton model to obtain the average number of $>$ 200 keV electrons per second and the power-law spectral index in each 1 minute interval, which are plotted in Figure \ref{fig:number_index}~b and \ref{fig:number_index}~c with $\pm1\sigma$ statistical uncertainties. We do not show data after 11:24:42 UT due to significant uncertainties in the estimate of the electron number and the spectral index.\\
   
Collisionless shocks with fast magnetosonic Mach numbers ($M_{\rm fm}$) greater than 2.7 are supercritical \citep{1995LNP...444..183M, 1998ISSIR...1..249S, 2016RPPh...79d6901M}. Supercritical shocks cannot be stabilized by Ohmic dissipation alone and they are able to accelerate particles efficiently to high energies. During the onset of the Fermi event, the area of the shock magnetically connected to the visible disk was supercritical and quasi-perpendicular. The quasi-perpendicular geometry resulted from the fact that magnetic connection occurred on the flank of the CME in a predominantly vertical magnetic field. The flank of the shock was moving at speeds in excess of 1000 km/s in quiet regions of the solar atmosphere with typical magnetosonic speeds of 300-400 km/s. In these regions a supercritical shock could form \citep{2017A&A...608A..43P}. These properties provided favorable conditions for the acceleration of electrons to high energy via Shock-Drift Acceleration (SDA) \citep{1983ApJ...267..837H, 1984JGR....89.8857W}. In this paper, we go one step further than \citet{2017A&A...608A..43P} and \citet{2019ApJ...876...80K}. We employ a test-particle simulation to compute the acceleration process of elections under these conditions.\\

Low in the corona, the CME shock propagated through both open and closed magnetic field lines. During the event of interest here, detailed modeling showed that the coronal shock was predominantly connected magnetically to the chromosphere by magnetic loops during the first hour of the eruption \citep{2017A&A...608A..43P}. Very few studies have investigated particle acceleration by shock waves along magnetic loops. Recent theoretical studies have investigated the effects of field-intensity gradient and field-line curvature on particle acceleration at shocks \citep{2006A&A...455..685S}. \citet{2009A&A...507L..21S} have computed the impact on the shock acceleration process of the curvature and expansion rate of open magnetic fields that form helmet streamers . Other studies \citep{2017ApJ...851...38K} have evaluated the efficiency of particle acceleration along streamer loops, where particles are mainly trapped and accelerated around the loop top. In this paper we study particle acceleration along loops that connect the expanding shock to the visible disk. In particular we study whether such geometries would allow particles accelerated at the shock to be mirrored back from the upstream footpoint on the visible disk (FP2 in Figure {\ref{fig:loc}}) by strong magnetic mirroring in the lower corona and chromosphere. We only consider precipitation at the footpoint in the shock's upstream in this paper. In a future study, we will study precipitation at the footpoint situated in the shock's downstream region (FP1 in Figure {\ref{fig:loc}}), as pointed out by \citet{2020SoPh..295...18G}.\\
   
In order to explain the origin of the hard X-ray emissions measured by Fermi GBM on 2014 September 1 we address a number of fundamental questions: (1) How efficiently are electrons accelerated by the mechanism of SDA during the event? (2) If SDA is insufficient, then what other mechanisms or combination of mechanisms could energise electrons to several hundred keVs or even tens of MeVs? (3) If we assume that electrons are indeed accelerated to high energies in the corona, what fraction of these particles can reach the solar surface to produce non-thermal emissions?\\
 
\subsection{Electron acceleration at coronal shocks} \label{sec:paracc}
    
Fast-mode shocks can be treated as moving magnetic mirrors because they are associated with a magnetic field compression. In the de Hoffmann-Teller (HT) frame, the plasma flow is along the magnetic field, and thus, the motional electric field is removed \citep{1950PhRv...80..692D}. In this frame, when a particle encounters a shock with a pitch angle such that it is outside of the loss cone of the shock, it is reflected rather than transmitted. The acceleration of reflected particles is due to reflection off a moving magnetic mirror, i.e. Fermi acceleration \citep{2001PASA...18..361B}. Comparison of observed events with models of particle acceleration at interplanetary shocks verifies that when the shock is quasi-perpendicular, SDA plays a key role in particle acceleration when the level of magnetic fluctuations is low and scattering is unimportant. Past studies have argued that SDA is responsible for the shock spikes, a strong enhancement in flux perpendicular to the magnetic field direction, measured predominantly at quasi-perpendicular interplanetary shocks \citep{1994ApJS...90..553E}. \citet{2009A&A...494..669M} applied this mechanism to shock waves produced during solar flares. They computed the production rate and the power of accelerated electrons in the flare region that compared favorably to RHESSI observations of the solar event on October 28, 2003. It is also expected that SDA is responsible for the occurrence of Type II solar radio bursts. \citet{2012JGRA..117.4106S} computed the properties of accelerated electron beams that can drive Langmuir waves and produce radio emissions.\\
    
By means of fully kinetic particle-in-cell (PIC) simulations, \citet{2014ApJ...794..153G} traced the evolution of the injected electrons and confirmed that the properties of the reflected electrons match the SDA predictions. This demonstrates that a fraction of electrons indeed participate in SDA.\\
    
Magnetic fluctuations can come from photospheric convection or nanoflaring. When magnetic fluctuations are introduced, particles can be scattered back to the shock multiple times, gaining more energy with each SDA cycle, which is known as the Diffusive-Shock Acceleration (DSA) process \citep{1988SSRv...48..195D}. We refer to SDA as simply the drift process, which is the principle acceleration mechanism in this paper. The drift process remains evident in DSA.\\
    
Under flare or shock conditions, at the onset of particle acceleration, several cycles of SDA can provide seed particles for the process of DSA. \citet{2018ApJ...869..182S} suggested that those seed particles are accelerated by the shock to produce the late phase $>$ 100 MeV $\gamma$-ray emission and contribute to the Solar Energetic Particles (SEPs). Each SDA cycle can lead to a significant energy gain of electrons at quasi-perpendicular shocks due to the high projected shock speeds. And overall, less shock encounters are required for particles to gain significant energy at quasi-perpendicular shocks. In this first theoretical study we focus on the relative role of SDA and DSA for the acceleration of electrons that propagate along magnetic loops traversed by the expanding shock wave.\\

\subsection{Electron transport in the corona} \label{sec:parttran}

To interpret Fermi observations we must determine under what conditions particles are able to overcome the strong mirroring from FP2 in Figure {\ref{fig:loc}} as they propagate toward the stronger magnetic fields that prevail near the solar surface. Indeed the mirror force will strongly limit the number of energetic electrons and protons that can hit the solar surface to produce hard X-rays and $\gamma$-rays, respectively.\\
    
\citet{1989ApJ...341..516H} and later \citet{2007ApJS..168..167M} discussed how turbulence in flare closed loops affect the ability for protons to reach the solar atmosphere in the presence of strong magnetic mirroring. The scattering of particles in pitch angle off magnetic irregularities that may exist in coronal loops could play an important role in counteracting the effect of the magnetic mirror by shifting particles into the loss cone of FP2 in Figure {\ref{fig:loc}}. These particles would then be able to precipitate. As already discussed, pitch-angle scattering of particles near the shock can also increase the number of shock crossings via the DSA process \citep[e.g.,][]{1978MNRAS.182..147B,1978MNRAS.182..443B,1982ApJ...255..716J,2014JSWSC...4A..08V,2018A&A...614A...4A}. In order to gain a more complete insight on the possible link between coronal shocks and the occurrence of high-energy radiation, it is therefore important to model the acceleration and transport of particles in terms of the pitch-angle scattering efficiency.\\
    
This paper then generally proceeds through a number of increasingly complex simulations. In Section 2, we first describe our shock-fitting technique and the method for obtaining shock-parameters along magnetic loops. In Section 3, we present our new Monte Carlo code that computes particle acceleration for a single interaction with the shock (one SDA cycle) and validate the computation by comparing our results with analytical SDA calculations (Appendix A). In Section 4, we set up the new model by using the shock properties in Section 2. We compute SDA induced by realistic shock and coronal conditions at different times and study the effect of mirroring of particles propagating toward the visible disk (Section 5). We then allow for multiple shock encounters along realistic magnetic loops (Section 6) and allow electrons to return from downstream (Section 7 and Appendix B). The next part of the paper investigates DSA (Section 8). We compare our one loop calculation with Fermi GBM observations in Section 9.\\

\section{Modeling the corona and the shock} \label{sec:modelcorshock}

In this section, we reconstruct the shock front and determine shock-parameters. We also present magnetic loops connecting the shock front to the solar disk visible from Earth by employing a 3-D global MHD model of the coronal magnetic field. Then the variability of shock and ambient plasma parameters along different magnetic loops is illustrated.\\

\subsection{Shock reconstruction and the background corona} \label{sec:shockrecon}

The 3-D evolution of the expanding shock wave is determined by fitting geometrical shapes to the multipoint observations provided by the EUV and white-light observations taken by the NASA Solar Terrestrial Relation Observatory (STEREO-A), the Solar Dynamic Observatory (SDO) spacecraft and SoHO. The technique is presented in \citet{2016ApJ...833...45R} and exploited in \citet{2017A&A...608A..43P}) and improved recently in \citet{2019ApJ...876...80K} in order to obtain reconstructions at higher cadence. The shock geometry is assumed to be an ellipsoidal structure. And the 3-D kinematic properties of the shock are discussed below.\\
   
To obtain the properties of the shock such as the Mach number, compression ratio, and its geometry, we must know the conditions upstream of the shock and thus determine the properties of the background corona into which the shock is propagating. We follow a similar procedure detailed in our past studies \citep{2016ApJ...833...45R, 2017A&A...608A..43P, 2019ApJ...876...80K} by using the coronal magnetic fields, plasma density and temperature obtained from a 3-D global MHD model developed and run by Predictive Sciences Inc. The Magneto-Hydrodynamic Around a Sphere Thermodynamic (MAST) model, uses magnetograms as boundary conditions at the Sun and solves the 3-D MHD equations in spherical coordinates from the upper chromosphere to the upper corona with an outer boundary set at 30 solar radii. The magnetogram used here is obtained by the heliospheric imager (HMI) onboard SDO on 2014 September 1 as input at the lower boundary \citep{2009ApJ...690..902L}. In order to simulate coronal properties accurately, past studies \citep{1988ApJ...325..442W, 2001ApJ...546..542L} have shown that energy exchanges that occur in the low corona must be solved for explicitly. This includes heating, thermal conduction along the magnetic field, and radiative losses. The MAST model does that and also includes the effects of Alfv\'{e}n waves on heating and pressure.\\
    
\begin{figure}[hbtp!]
    \plotone{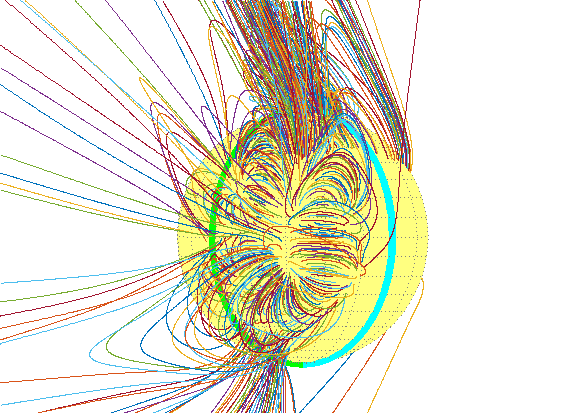}
    \caption{All field lines that start from the visible disk and intersect the shock at some point. The East Limb of the Sun as visible from Earth is indicated by a light green arc. The cyan arc shows the Central Meridian of the Sun.
    \label{fig:allfieldlines}}
\end{figure}
    
\begin{figure}[hbtp!]
    \plotone{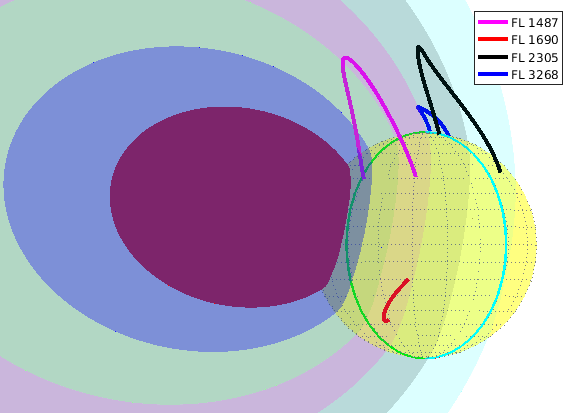}
    \caption{4 field lines of particular interest. The ellipsoids show the shock locations at 11:06, 11:10, 11:14, 11:18, 11:22 and 11:26 UT. The East Limb of the Sun as visible from Earth is indicated by a light green arc. The cyan arc shows the Central Meridian of the Sun.
    \label{fig:4lines}}
\end{figure}
    
The 3-D speed of the shock and the conditions upstream can be used to derive Mach numbers at the shock and by solving the Rankine-Hugoniot relation \citep[see e.g.,][]{1982soma.book.....P, 2007ShWav..16..477S} the gas and magnetic compression ratios can also be obtained. From the 3-D modeling of the shock and the tracing of magnetic field lines, we can infer how the shock connects magnetically to the visible disk and thus interpret how particles accelerated at the shock may stream down to the lower corona. This was discussed in some detail for that event by \citet{2017A&A...608A..43P}. We trace a large number of these magnetic field lines in Figure \ref{fig:allfieldlines}. It was found in our previous study that the onset of the high-energy emissions in hard X-rays and $\gamma$-rays measured by Fermi occurred when the shock connected to the visible disk. Moreover, \citet{2017A&A...608A..43P} showed the shock was initially connected to magnetic loops. As particle acceleration and transport along such loops has not been studied in great detail, this is a point of focus in this article.\\

\subsection{Shock parameters along selected magnetic loops}
\label{sec:shockpar}

We model particle acceleration along a single magnetic loop to demonstrate the important physical processes at play in the corona that accelerate electrons to relativistic energies. We begin by illustrating the large variability of coronal loop properties (e.g. length and orientation) by selecting four representative field lines that intersect different regions of the shock flank. These lines are shown in Figure \ref{fig:4lines} together with the expanding shock. The plasma temperature $T$, magnetic field magnitude $B$, plasma density $n$ and the value of the magnetic field inclination to the shock-normal $\theta_{\rm Bn}$ are then determined at the shock along these 4 field lines as a function of time as the shock expands.\\
    
\begin{figure*}[hbtp!]
    \epsscale{1.2}
    \plotone{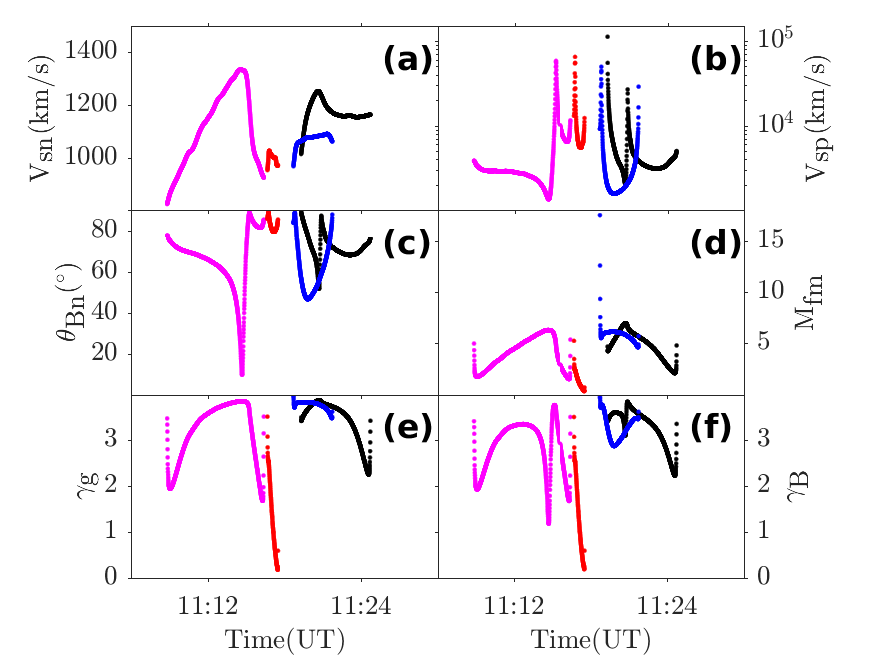}
    \caption{Shock parameters along the 4 selected field lines: FL 1487 (magenta), FL 1690 (red), FL 2305 (black) and FL 3268 (blue). Shock normal speed $V_{\rm sn}$ (Panel a), projected shock speed $V_{\rm sp}$ (Panel b), the value of the magnetic field inclination to the shock-normal $\theta_{\rm Bn}$ (Panel c), fast-mode Mach number $M_{\rm fm}$ (Panel d), gas compression ratio $\gamma_{\rm g}$ (Panel e) and magnetic compression ratio $\gamma_{\rm B}$ (Panel f).
    \label{fig:subplot1}}
\end{figure*}
    
\begin{figure}[hbtp!]
    \epsscale{1.25}
    \plotone{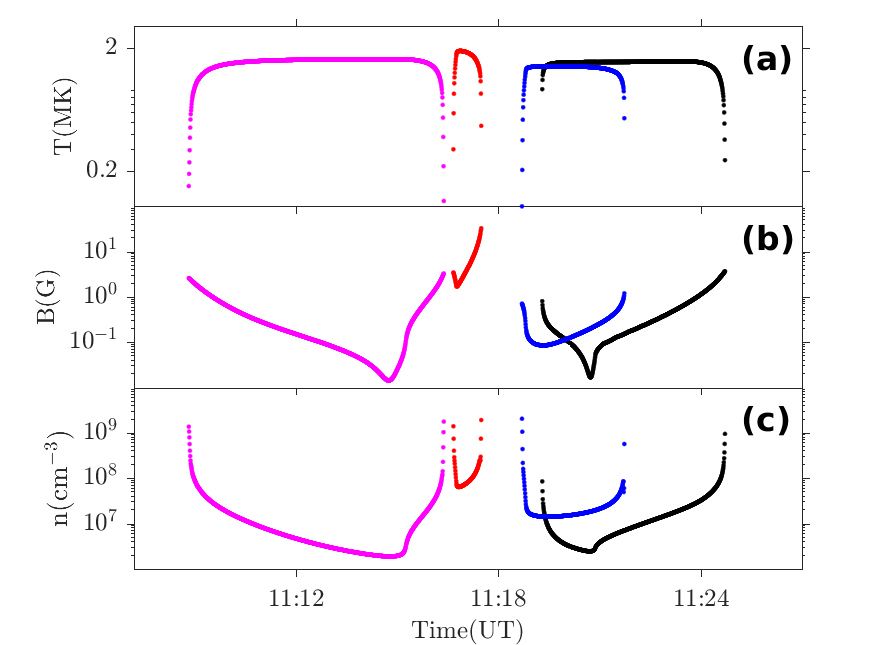}
    \caption{Ambient plasma parameters along the 4 selected field lines: FL 1487 (magenta), FL 1690 (red), FL 2305 (black) and FL 3268 (blue). Plasma temperature (Panel a), magnetic field magnitude (Panel b) and plasma density (Panel c).
    \label{fig:subplot2}}
\end{figure}
    
The finite resolution of both the shock reconstruction technique and the background coronal model leads to fluctuations in the shock parameters derived along specific field lines. Such fluctuations can create spurious effects that must be avoided in order to model accurately the acceleration and transport of particles. For instance in \citet{2018A&A...614A...4A}, analytical expressions are used to fit the shock parameters derived by \citet{2016ApJ...833...45R}. Here we take a slightly different approach to derive the shock parameters.\\
    
The initial step to derive shock speed is similar to \citet{2016ApJ...833...45R}. We consider a set of 200$\times$200 grid points distributed over the shock ellipsoid. To compute the shock normal speed $V_{\rm sn}$, we first find the grid point $P$ on the ellipsoid at time $t$ that is closest to the shock-field line intersection at time $t$, whose index is $(i,j)$. We then find the grid point $P'$ with index $(i,j)$ on the ellipsoid at time $t-\delta t$, where $\delta t$ varies between 0.5 second and 1.5 second. Next, we compute the distance between $P$ and $P'$, which we divide by the time interval $\delta t$. The speed of the propagating front along the field line is $V_{\rm sp}= V_{\rm sn}/\cos\theta_{\rm Bn}$. The fast-mode speed is then obtained by:
\begin{eqnarray}
V_{\rm fm}=\sqrt{\frac{1}{2}[V^2_{\rm A}+C^2_{\rm S}+\sqrt{(V^2_{\rm A}+C^2_{\rm S})^2-4V^2_{\rm A}C^2_{\rm S}\cos^2\theta_{\rm Bn}}]}\nonumber \\
\end{eqnarray}
where $V_{\rm A}$ is the Alfv\'{e}n speed and $C_{\rm S}$ is the sound speed. The fast magneto-sonic number $M_{\rm fm}$ is defined as
\begin{equation}
M_{\rm fm}=\frac{V_{\rm sn}-{\bf V}_{\rm w}\cdot{\bf n}}{V_{\rm fm}}
\end{equation}
where $\bf n$ is the local shock-normal vector and ${\bf V}_{\rm w}$ is the solar wind bulk velocity. Once the parameters of the evolving shock are determined along the four selected field lines, these parameters are spline interpolated at a cadence of $\delta t=0.5$ second to produce a sequence of regularly time-spaced ellipsoids.\\

Shock parameters and ambient plasma parameters along the four field lines are shown in Figures \ref{fig:subplot1} and \ref{fig:subplot2}. As shown in Figure \ref{fig:subplot2}~a, the temperature of the magnetic loops drops to chromospheric values at the footpoints, and increases radially outward to coronal values. Figure \ref{fig:subplot1}~a confirms that for most of its propagation through the loop, the shock normal speed exceeds 1000 km/s. During part of its propagation along the loops the shock exhibits a quasi-perpendicular geometry with values of $\theta_{\rm Bn}$ exceeding 80$^\circ$ (Figure \ref{fig:subplot1}~c) with very high speeds projected along the field line (Figure \ref{fig:subplot1}~b). The typical $M_{\rm fm}$ exceeds 2.7 meaning that the shock is supercritical (Figure \ref{fig:subplot1}~d). To compute the gas compression ratio $\gamma_{\rm g}$ and the magnetic compression ratio $\gamma_{\rm B}$, we follow the recipe presented in \citet{2015ApJ...809..111L}. \citet{2015ApJ...809..111L} defines the thermal ratio
\begin{eqnarray}
    \zeta_{\rm TV}\equiv\frac{2k_{\rm B}T}{m_0(V_{\rm sn}-{\bf V}_{\rm w}\cdot{\bf n})^2},
\end{eqnarray}
where $k_{\rm B}$ is the Boltzmann constant, $T$ is the temperature and $m_0$ is the rest mass of the particle (cf Eq. (4) in \citet{2015ApJ...809..111L}). \citet{2015ApJ...809..111L} also defines the ratio of the perpendicular component of the flow kinetic energy over that of the magnetic energy, 
\begin{eqnarray}
    \zeta_{\rm VB\perp}\equiv\frac{(V_{\rm sn}-{\bf V}_{\rm w}\cdot{\bf n})^2}{V^2_{\rm A}\cos^2{\theta_{\rm Bn}}},
\end{eqnarray}
\begin{eqnarray}
    \zeta_{\rm VB}\equiv\frac{(V_{\rm sn}-{\bf V}_{\rm w}\cdot{\bf n})^2}{V^2_{\rm A}}
\end{eqnarray}
(cf Eq. (23b) in \citet{2015ApJ...809..111L}). The gas compression ratio $\gamma_{\rm g}$ is given by the solution of the cubic trinomial, 
\begin{eqnarray}
    \Pi_3(\gamma_{\rm g})\equiv\sum_{k=0}^3q_{k}\gamma^k_{\rm g}=0
\end{eqnarray}
(cf Eq. (24a) in \citet{2015ApJ...809..111L}), where
\begin{eqnarray}
    q_0=-(1+d)\zeta^2_{\rm VB},
\end{eqnarray}
\begin{eqnarray}
    q_1=\left\{\left(1+\frac{3}{2}d\right)\cos^2{\theta_{\rm Bn}}+\left(1+\frac{1}{2}d\right)\right.\nonumber \\
    \left.+\left[1+\left(1+\frac{1}{2}d\right)\zeta_{\rm TV}\right]\zeta_{\rm VB}\right\}\zeta_{\rm VB},
\end{eqnarray}
\begin{eqnarray}
    q_2=-\left\{\left[\left(d+1\right)+\left(1+\frac{1}{2}d\right)(1+2\zeta_{\rm TV})\zeta_{\rm VB}\right]\right.\nonumber\\
    \left.\times\cos^2{\theta_{\rm Bn}}+\left(1-\frac{1}{2}d\right)\zeta_{\rm VB}\right\},
\end{eqnarray}
\begin{eqnarray}
    q_3=\left[1+\left(1+\frac{1}{2}d\right)\zeta_{\rm TV}\cos^2{\theta_{\rm Bn}}\right]\cos^2{\theta_{\rm Bn}},
\end{eqnarray}
and $d$ indicates the degrees of freedom (cf Eq. (24c) in \citet{2015ApJ...809..111L}). For the adiabatic case here, $d=3$.\\

According to \citet{1992nrfa.book.....P}, the three solutions of $\Pi_3(\gamma_{\rm g})$ are given by
\begin{eqnarray}
    \gamma_{\rm g}=-\frac{1}{3q_3}\cdot\left(q_2+hY+h^*\frac{\Delta_0}{Y}\right)
\end{eqnarray}
(cf Eq. (25a) in \citet{2015ApJ...809..111L}) where
\begin{eqnarray}
    Y\equiv\sqrt[3]{\frac{\Delta_1+\sqrt{\Delta^2_1-4\Delta^3_0}}{2}},
\end{eqnarray}
\begin{eqnarray}
    \Delta_0{\equiv}q_2^2-3q_1q_3,
\end{eqnarray}
\begin{eqnarray}
    \Delta_1{\equiv}2q_2^3-9q_1q_2q_3+27q_0q_3^2
\end{eqnarray}
(cf Eq. (25b) in \citet{2015ApJ...809..111L}) and $h$ represents any of the (complex) cubic roots of unity
\begin{eqnarray}
    h=1, \left(-1+i\sqrt3\right)/2, \left(-1-i\sqrt3\right)/2,
\end{eqnarray}
($h^*$ is the conjugate of $h$) (cf Eq. (25c) in \citet{2015ApJ...809..111L}). We limit solutions to fast-mode shocks. The magnetic compression ratio $\gamma_{\rm B}$ is derived from Eq. (22c) in \citet{2015ApJ...809..111L}.
\begin{eqnarray}
    \gamma_{\rm B}=\cos{\theta_{\rm Bn}}\cdot\sqrt{1+\gamma_{\rm g}^2(\xi+1)^2\cdot\tan^2{\theta_{\rm Bn}}}
\end{eqnarray}
where
\begin{eqnarray}
    \xi\equiv\frac{1-\gamma^{-1}_{\rm g}}{\gamma^{-1}_{\rm g}\zeta_{\rm VB\perp}-1}
\end{eqnarray}
(cf Eq. (23a) in \citet{2015ApJ...809..111L}).\\

Both the gas (Figure \ref{fig:subplot1}~e) and magnetic (Figure \ref{fig:subplot1}~f) compression ratios often exceed 2. Such a fast supercritical and quasi-perpendicular shock provides favorable conditions for SDA to occur along these loops.\\

Since these parameters are spline interpolated at a cadence of $\delta t=0.5$ second, the time difference between neighboring data points is 0.5 second in Figures \ref{fig:subplot1} and \ref{fig:subplot2}. At the footpoints of loops, the magnetic fields are almost radial, which results in large shock angles and large $V_{\rm sp}$. Thus, at the footpoints of loops, the difference in heliocentric distances between neighboring data points is large. Therefore, in Figure \ref{fig:subplot2}, the plasma temperature shows an abrupt decrease at the footpoints, while the plasma density shows an abrupt increase at the footpoints. And consequently, the fast magneto-sonic number $M_{\rm fm}$, the gas compression ratio $\gamma_{\rm g}$ and the magnetic compression ratio $\gamma_{\rm B}$ increase abruptly at the footpoints.\\
    
From the four field lines shown in Figure \ref{fig:4lines}, the extended loops FL 1487 and FL 2305 were well suited to study the bouncing of particles between converging magnetic mirrors imposed by the moving shock front and the strong magnetic fields at the footpoint of the loop, an effect our new model was designed to address. FL 1487 was actually connected at one footpoint very close to the $\gamma$-ray source and we have started simulating protons and electrons on FL 1487, this ongoing work will be presented in a future paper. The present paper is focused on the hard X-ray source and FL 2305 was particularly interesting to the hard X-ray problem. The time interval over which the shock intersects this magnetic field line is shaded in yellow in Figure \ref{fig:number_index}. The shock propagated through FL 2305 after the hard X-rays derived from soft X-ray measurements of the on-disk flare viewed from MESSENGER had ceased. There appears to be no softening of the GBM hard X-ray spectra during that phase and our hypothesis was that hard X-rays could be further produced by the shock during late connection to loops such as FL 2305. In future work we plan to produce a full synthetic time-dependent spectrum of hard X-ray emission integrating the contribution of all field lines.\\

\section{SDA computed with a Monte Carlo approach} \label{sec:shockdrift}

We have developed a Monte Carlo code which is capable of simulating SDA. We compute the change in pitch angle and speed of an electron reflected at the moving shock in the HT frame, in which the convective electric field disappears. Electron velocities parallel and perpendicular to the magnetic field before the reflection are $v_{\rm i,\parallel}$ and $v_{\rm i,\perp}$. Dividing them by the speed of light $c$, we obtain $\beta_{\rm i,\parallel}=v_{\rm i,\parallel}/c$ and $\beta_{\rm i,\perp}=v_{\rm i,\perp}/c$. Electron velocities parallel and perpendicular to the magnetic field after the reflection are $v_{\rm r,\parallel}$ and $v_{\rm r,\perp}$. Similarly, $\beta_{\rm r,\parallel}=v_{\rm r,\parallel}/c$ and $\beta_{\rm r,\perp}=v_{\rm r,\perp}/c$. $\beta_{\rm i,\parallel}$ and $\beta_{\rm i,\perp}$ are related to $\beta_{\rm r,\parallel}$ and $\beta_{\rm r,\perp}$ by
\begin{eqnarray}
\beta_{\rm i,\parallel}=\frac{2\beta_{\rm s}-\beta_{\rm r,\parallel}(1+\beta^2_{\rm s})}{1-2\beta_{\rm r,\parallel}\beta_{\rm s}+\beta^2_{\rm s}}
\label{7}
\end{eqnarray}
(cf Eq. (1) in \citet{2009A&A...494..669M}) and
\begin{eqnarray}
\beta_{\rm i,\perp}=\frac{(1-\beta^2_{\rm s})}{1-2\beta_{\rm r,\parallel}\beta_{\rm s}+\beta^2_{\rm s}}\cdot \beta_{\rm r,\perp}
\label{8}
\end{eqnarray}
where $\beta_{\rm s}=V_{\rm sp}/c$ (cf Eq. (2) in \citet{2009A&A...494..669M}). Here $\beta_{\rm i,\parallel}$, $\beta_{\rm i,\perp}$, $\beta_{\rm r,\parallel}$ and $\beta_{\rm r,\perp}$ are all in the upstream plasma frame.\\

We assume $\alpha$ is the electron pitch angle in the HT frame and $B_1$ and $B_2$ are the upstream and downstream magnetic field strengths. The electron will be transmitted to downstream if 
    \begin{equation}
    \sin^2\alpha\le\sin^2\alpha_{\rm c}=B_1/B_2=1/\gamma_{\rm B}.
    \end{equation}
 Otherwise, it will be reflected at the shock. This criterion is equivalent to the reflection conditions
    \begin{equation}
    \beta_{\rm i,\parallel}\le\beta_{\rm s}
    \end{equation}
(cf Eq. (3) in \citet{2009A&A...494..669M}) and
    \begin{equation}
    \beta_{\rm i,\perp}\ge\frac{\tan \alpha_{\rm c}}{\sqrt{1-\beta^2_{\rm s}}}\cdot(\beta_{\rm s}-\beta_{\rm i,\parallel})\\
    \end{equation}
(cf Eq. (4) in \citet{2009A&A...494..669M}). After the reflection, the electron gets a boost to its parallel speed, whereas its pitch angle decreases. Eqs. (21) and (22) provide the basic conditions for electron-shock interactions, which we implement in the Monte Carlo code.\\

In Appendix A, we test our Monte Carlo simulation of a single interaction of electrons with the shock by comparing our simulated distribution functions with those derived analytically \citep{2009A&A...494..669M}.\\

\begin{deluxetable*}{c|cccccc}[hbtp!]
\tablecaption{Shock properties at locations I, II, III, IV, V and VI \label{tab:prop}}
\tablehead{
\colhead{Shock} & \colhead{Shock} & \colhead{Projected} & \colhead{Shock} & \colhead{Fast-mode} & \colhead{Gas} & \colhead{Magnetic} \\
\colhead{location} & \colhead{normal} & \colhead{shock} & \colhead{normal} & \colhead{Mach} & \colhead{compression} & \colhead{compression} \\
\colhead{} & \colhead{speed} & \colhead{speed} & \colhead{angle} & \colhead{number} & \colhead{ratio} & \colhead{ratio} \\
\colhead{} & \colhead{(km/s)} & \colhead{(km/s)} & \colhead{($^\circ$)} & \colhead{} & \colhead{} & \colhead{} 
}
\startdata
I & $1.02\times 10^{3}$ & $5.48\times 10^{4}$ & 88.9 & 3.77 & 3.44 & 3.44 \\
II & $1.10\times 10^{3}$ & $9.83\times 10^{3}$ & 83.6 & 4.08 & 3.56 & 3.54 \\
III & $1.25\times 10^{3}$ & $2.38\times 10^{3}$ & 58.4 & 6.05 & 3.88 & 3.34 \\
IV & $1.16\times 10^{3}$ & $3.32\times 10^{3}$ & 69.5 & 4.47 & 3.65 & 3.46 \\
V & $1.16\times 10^{3}$ & $4.14\times 10^{3}$ & 73.7 & 1.87 & 2.30 & 2.27 \\
VI & $1.16\times 10^{3}$ & $4.81\times 10^{3}$ & 76.0 & 2.48 & 2.77 & 2.73 \\
\enddata
\end{deluxetable*}

\section{Numerical Setup for a magnetic flux tube} \label{sec:setup}

    \begin{figure*}[hbtp!]
    \epsscale{1.15}
    \plotone{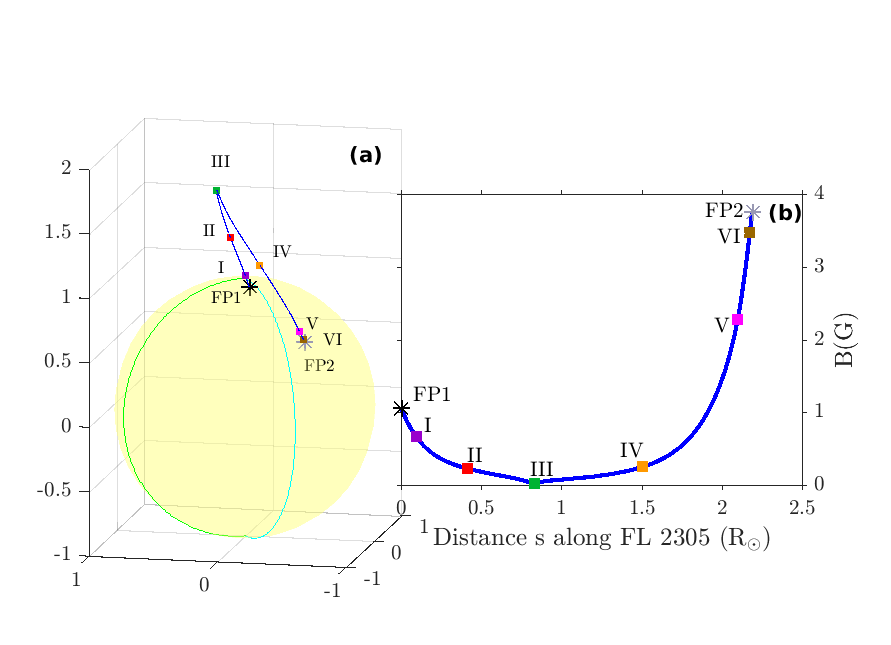}
    \caption{Shock locations I (purple), II (red), III (green), IV (orange), V (magenta) and VI (brown) on the loop FL 2305 where the shock was fixed for single interaction or forced to stop for multiple interactions (Panel a). The East Limb of the Sun as visible from Earth is indicated by a light green arc. The cyan arc shows the Central Meridian of the Sun. Magnetic field strength B vs. distance s along FL 2305 (Panel b). Footpoint 1 (FP1, i.e. the downstream footpoint) and footpoint 2 (FP2, i.e. the upstream footpoint) are indicated by a black star and a gray star respectively. Both FP1 and FP2 are on the visible disk.
    \label{fig:loc}}
    \end{figure*}
    
    \begin{figure*}[hbtp!]
    \centering
    \includegraphics[width=1.5\linewidth]{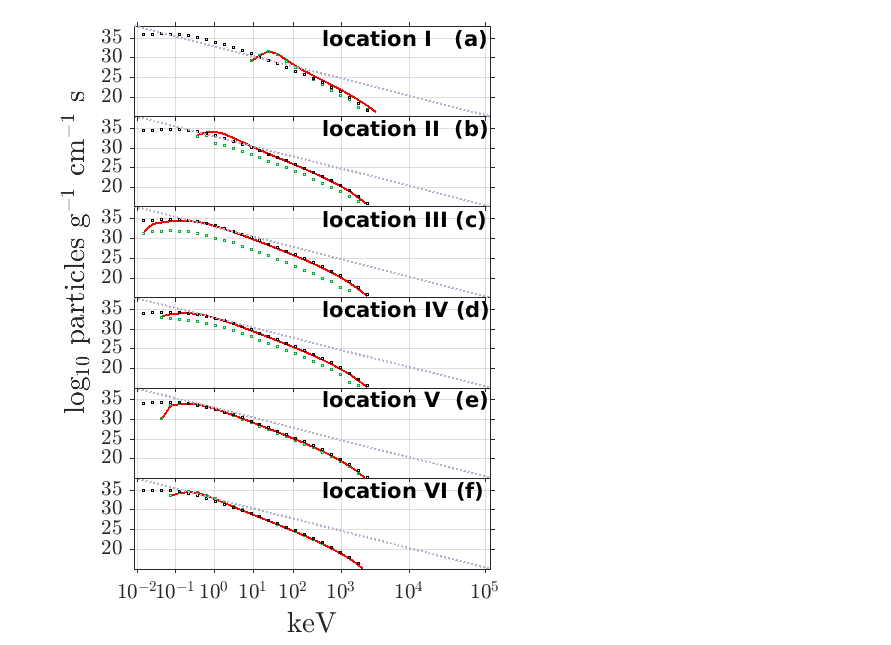}
    \caption{Simulated electron spectra for which only a single interaction with the shock is allowed. The spectra of accelerated electrons are plotted as red curves. Black squares show the initial spectra. The spectra of precipitating electrons are presented as green squares. The dotted gray line in each panel is a power-law with spectral index of 5 in momentum space.
    \label{fig:single}}
    \end{figure*}
    
    \begin{figure}[hbtp!]
    \centering
    \includegraphics[width=1.0\linewidth]{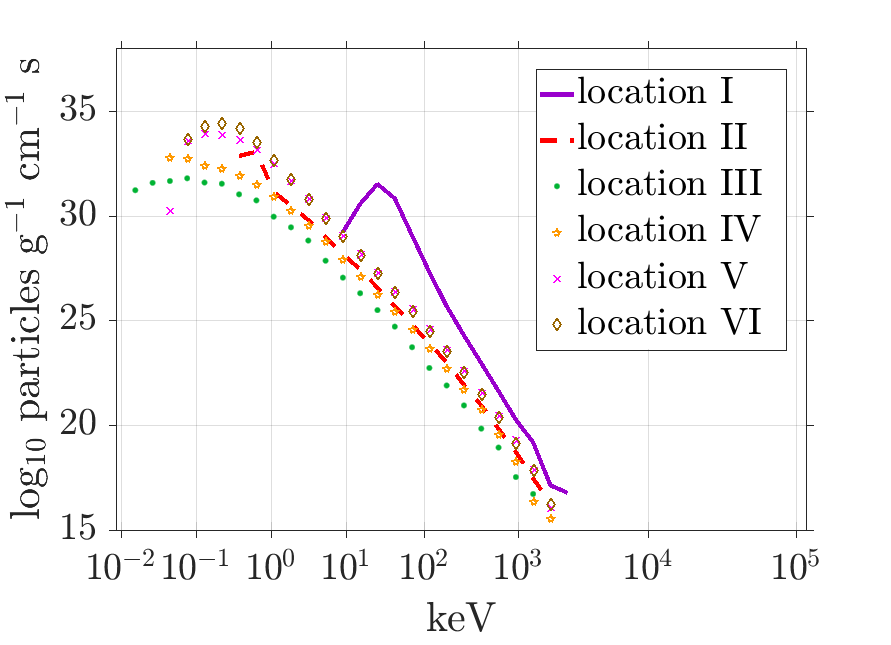}
    \caption{Simulated spectra of precipitating electrons for which only a single interaction with the shock is allowed: shock location I (solid purple curve), shock location II (dashed red curve), shock location III (green dots), shock location IV (orange pentagons), shock location V (magenta crosses) and shock location VI (brown diamonds).
    \label{fig:singlepreci}}
    \end{figure}
    
We now present the results of several numerical experiments illustrating the high variability and dynamic nature of particle acceleration that unfolds when shock waves propagate along a single loop. As already stated, we focus here on FL 2305. We run our Monte Carlo code along the magnetic loop by employing the plasma properties derived from the 3-D MHD MAST model. The shock properties, the injected particle population and the magnetic field lines along which the particles are propagating are all based on MAST. As illustrated in Figures \ref{fig:subplot1} and \ref{fig:subplot2}, we can define the time evolution of the shock and the properties of the plasma immediately upstream of the shock from the MAST model.\\

A single magnetic flux tube is assumed to be surrounding FL 2305. Electrons are considered confined within this magnetic flux tube. The cross-sectional area of the magnetic flux tube $A(s)$ scales as $B(s)^{-1}$, where $B(s)$ is the magnetic flux density. Both $A(s)$ and $B(s)$ are functions of the position along FL 2305 $s$. The length of FL 2305 $s_0 \approx 2.19 R_\odot$, where $R_\odot$ is the solar radius. The area of the solar surface in the shock's upstream covered by this magnetic flux tube is assumed to be 1 cm$^2$. The choice of 1 cm$^2$ is made to make the magnetic flux tube relatively thin. It has no impact on the total number of precipitating electrons over the entire visible disk. The cross-sectional area of the magnetic flux tube at FP2, $A(s_0)$, is then formulated as $A(s_0)=\sin \psi \cdot 1$ cm$^2$, where $\psi$ is the angle between the magnetic field line and the solar surface at FP2.\\

We introduce an injection time step $\Delta t_{\rm inj}$. The number of physical electrons $N(s)$ in a small spatial volume $A(s)V_{\rm sp}(s)\Delta t_{\rm inj}$ ahead of the shock is computed using plasma density $n(s)$ from MAST 
\begin{eqnarray}
N(s)=n(s)A(s)V_{\rm sp}(s)\Delta t_{\rm inj}.
\end{eqnarray}
We assume 117 logarithmic bins in the direction of momentum with a minimum momentum $p_{\rm min}$ = 1.7$\times$10$^{-19}$ g$\cdot$cm/s which corresponds to a minimum energy of $E_{\rm min}$ = 0.01 keV. The maximum momentum $p_{\rm max}$ = 5.4$\times$10$^{-15}$ g$\cdot$cm/s corresponds to a maximum energy $E_{\rm max}$ = 1$\times$10$^5$ keV. Since the extremely high energy electrons would not be sustained in the loops without a strong acceleration mechanism that can overcome the continuous precipitation, in all the following simulations in this paper we assume that the initial electrons follow an e-folding spectrum built from a $\kappa$-distribution of $\kappa$ = 3.5 with an exponential cutoff at $p_{\rm cut}$ = 1.8$\times$10$^{-17}$ g$\cdot$cm/s that corresponds to the cutoff energy $E_{\rm cut}$ = 100 keV. A value $\kappa$ = 3.5 is reasonable in the corona \citep{1999JGR...10417021P}. We have also verified that those electrons in the tail of the $\kappa$-distribution do not contribute much.\\

Physical electrons in the momentum bin $p_{\rm i}$ are represented by pseudo-particles assigned to the momentum bin $p_{\rm i}$. In a small spatial volume $A(s)V_{\rm sp}(s)\Delta t_{\rm inj}$, and for a momentum bin at $p_{\rm i}$, we inject $N_{\rm i}$ pseudo-particles where $N_{\rm i}=N_0\cdot\exp(-p_{\rm i}/p_{\rm cut})$. $N_0=50000$ for simulations that only a single interaction with the shock is allowed; while $N_0=2000$ for simulations that allow for multiple shock interactions per electron. The number of physical electrons in the momentum bin $p_{\rm i}$ is
\begin{eqnarray}
N_{\rm e}(p_{\rm i})=\int_{p_{\rm i-1/2}}^{p_{\rm i+1/2}}\frac{dN(s)}{dp}dp.
\end{eqnarray}
We divide $N_{\rm e}(p_{\rm i})$ by $N_{\rm i}$ and obtain the weight $W_{\rm i}$ of the injected pseudo-particle assigned to the momentum bin $p_{\rm i}$
\begin{eqnarray}
W_{\rm i}=\frac{N_{\rm e}(p_{\rm i})}{N_{\rm i}}.
\end{eqnarray}
In this paper, by ``inject" we mean that we place $N_{\rm i}$ pseudo-particles at the shock.\\ 

\section{Single Interaction} \label{sec:shockdrifttran}

We begin our numerical study by illustrating the efficiency of particle acceleration at different shock locations along a single magnetic loop (FL 2305) and do not model yet that a particle may encounter the shock multiple times. Thus, we compute particle acceleration assuming that each pseudo-particle can interact at most once with the shock (one SDA cycle) and the shock is fixed at location I, II, III ... or VI shown in Figure {\ref{fig:loc}}. We have calculated the electron spectra for these six runs. The shock reaches locations I-VI at 11:19:19, 11:19:31, 11:20:45, 11:22:21, 11:24:29 and 11:24:42 UT respectively. In Figure \ref{fig:number_index} we plot times when FL 2305 intersects the shock as yellow shaded areas and denote the times at which the shock crosses locations I, II, III ... VI on FL 2305 with vertical lines. Table \ref{tab:prop} gives the shock properties at all these six shock locations. Our simulation here consists of two steps.\\ 

Step 1: The shock is fixed at location I, II, III ... or VI shown in Figure {\ref{fig:loc}}. We inject pseudo-particles at each shock location with $N_0$ = 50000. Then we perform shock interaction routines: If a pseudo-particle satisfies the reflection conditions Eqs. (21) and (22), it is reflected at the shock. This pseudo-particle is labelled as an accelerated pseudo-particle. If a pseudo-particle fulfills Eq. (21), but it is inside the loss cone of the shock, it is absorbed, i.e. removed from the simulation. If a pseudo-particle does not satisfy Eq. (21), it does not interact with the shock. This pseudo-particle is labelled as an escaping background pseudo-particle. Accelerated and escaping background pseudo-particles are collected. And based on the momenta, pitch-angles and weights of these pseudo-particles, we plot the logarithm of physical electron number per momentum as a function of the logarithm of momentum using a base 10 logarithmic scale on the vertical and horizontal axes. Then we only place $10^{-2}$, $10^{-1}$...$10^{5}$ keV along the horizontal axis. The horizontal axis still represents the logarithm of momentum. Thus $10^{-2}$, $10^{-1}$...$10^{5}$ keV are not evenly spaced along the horizontal axis. This is how we compute electron spectra in all the following figures as well. From 117 logarithmic bins in the direction of momentum $\Delta p_1$, $\Delta p_2$, $\Delta p_3$ ... $\Delta p_{117}$, we select a subset of 39 bins $\Delta p_3$, $\Delta p_6$, $\Delta p_9$ ... $\Delta p_{117}$. For all the spectra, we only plot the values for these 39 bins. The red curves in Figure {\ref{fig:single}} show the accelerated electron momentum spectra at each shock location. The initial spectra are shown as black squares for comparison. Among all the initial electrons, those with parallel speed that is higher than the projected speed of the shock along the magnetic field do not interact with the shock. We call them escaping background electrons.\\ 

Step 2: The shock is still fixed at location I, II, III ... or VI. The simulation is performed in a fixed spatial simulation box in the upstream region of the shock. The box is bounded by the shock front and the visible disk at FP2 in Figure {\ref{fig:loc}}. All the accelerated and escaping background pseudo-particles collected in Step 1 are released at the shock. They then propagate from the shock toward the solar surface in the absence of shock interaction. The propagation of the pseudo-particle can be described by guiding-center’s motion along the magnetic field and pitch-angle focusing: 
\begin{eqnarray}
\frac{{\rm d}s}{{\rm d}z}=\mu
\end{eqnarray}
and
\begin{eqnarray}
\frac{{\rm d}\mu}{{\rm d}z}=-\frac{(1-\mu^2)}{2B(s)}\cdot\frac{\partial B(s)}{\partial s},
\end{eqnarray}
where $\mu$ is the pitch-angle cosine and $z=vt$ with $v$ being the speed of the pseudo-particle. Eqs. (26) and (27) are solved with an Euler scheme. Pseudo-particles that cross the boundary at FP2 in Figure {\ref{fig:loc}} are absorbed, i.e. precipitated. These pseudo-particles are labelled as precipitating pseudo-particles. Based on their momenta, pitch-angles and weights, we compute the spectra of precipitating electrons and plot them as green squares in Figure {\ref{fig:single}}.\\ 

Since only a single shock encounter is allowed, the acceleration is overall limited, and the spectra of accelerated electrons (red curves) are very close to the initial spectra (black squares) in Figure {\ref{fig:single}}. The accelerated electron velocity component parallel to the magnetic field has to be greater than the shock speed projected to the field line. Thus, the accelerated electron velocity exceeds the shock speed projected to the field line which varies between different locations. Thus, the low-energy cutoff of the red curve in Figure {\ref{fig:single}} varies between the different spectra. At location I, the shock is nearly perpendicular ($\theta_{\rm Bn}\sim89^\circ$), therefore, the shock speed projected to the magnetic field line is much higher (see Table \ref{tab:prop}). Thus, the shock at location I accelerates electrons to much higher energies than the shock at any other locations, with a peak occurring near 26 keV. At location II, $\theta_{\rm Bn}\sim84^\circ$, the shock becomes a little less efficient. The red curve almost overlaps the black squares at shock location III, which means the shock does not really accelerate electrons much. This is due to a much smaller $\theta_{\rm Bn}$, which is about $58^\circ$. There is a little more acceleration at shock location VI because $\theta_{\rm Bn}$ increases from shock location III to shock location VI. Figure {\ref{fig:singlepreci}} shows the varying precipitation for different injected locations. The significant particle acceleration computed near shock location I occurs soon after the shock first intersects the loop near one footpoint on the visible disk (FP1 in Figure {\ref{fig:loc}}). Precipitation of the electrons at FP2 in Figure {\ref{fig:loc}} is also very high since the two footpoints share similar magnetic field strengths and therefore particles experience little mirroring from FP2 during their propagation. One can see that the closer to the loop top the acceleration occurs, the fewer electrons can precipitate due to the large magnetic field ratio between FP2 and the top of the magnetic loop that imposes a stronger mirror force on the propagating particles.\\

\section{Multiple Interactions} \label{sec:multinter}

Some electrons interacting once with the shock can be mirrored back during their propagation to FP2 and be naturally brought back to the shock. This could force a second shock interaction and further acceleration of electrons, which was one of the mechanisms that lead to multiple interactions with the shock studied by \citet{2006A&A...455..685S}. Some electrons interacting once with the shock can also be caught up with by the shock when the shock turns more and more perpendicular. These electrons receive a boost to their parallel speed even without the mirror force, which was described as the HT resonance effect in \citet{2006A&A...455..685S} as well. In this section we allow electrons to interact with the shock multiple times to illustrate the dramatic increase in energy experienced by these electrons.\\

We now compute particle acceleration and transport by allowing multiple shock interactions per electron. We have calculated the electron spectra for three runs. Run 1: the shock travels from the first shock-field line intersection to location II. Run 2: the shock travels from the first shock-field line intersection to location III. Run 3: the shock travels from the first shock-field line intersection to location VI. In this section as well as Sections 7, for each run, our simulation consists of two steps.\\ 

\begin{figure*}[hbtp!]
    \centering
    \includegraphics[width=0.75\linewidth]{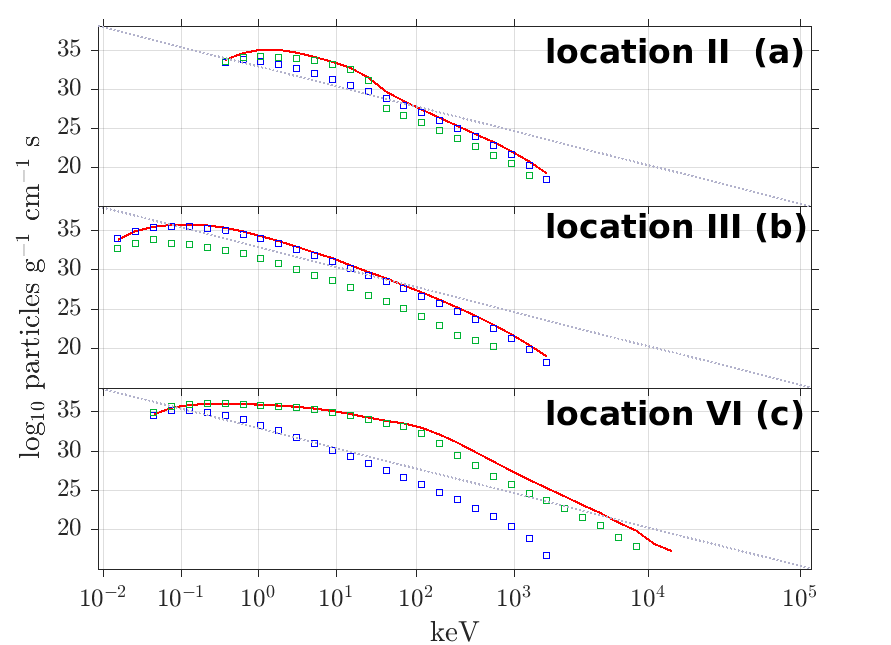}
    \caption{In the same format as Figure {\ref{fig:single}}, we show particle energy spectra obtained upstream of the shock for simulations that allow for multiple shock interactions per particle but do not allow particles to be scattered downstream of the shock to return to the shock. The shown spectra are for shock locations II (a), III (b) or VI (c) only because they exhibit the most interesting features. As in Figure {\ref{fig:single}}, they include accelerated electrons (red curves), the escaping background electrons (blue squares) and the electrons that precipitate (green squares). The dotted gray line in each panel is a power-law with spectral index of 5 in momentum space. 
    \label{fig:noreturn}}
    \end{figure*}

Step 1:  The simulation is performed in a shrinking spatial simulation box in the upstream region of the shock. The box is bounded by the moving shock front propagating along the magnetic loop and the visible disk at FP2 in Figure {\ref{fig:loc}}. We inject $N_0$ = 2000 new pseudo-particles every 0.5 second, i.e. $\Delta t_{\rm inj}=0.5$ second, as a means of simulating the number of ambient electrons in the loop as the shock moves along. For each pseudo-particle, Eqs. (26) and (27) are solved with a fourth-order Runge-Kutta scheme \citep{1895soma.book.....R,1901soma.book.....K}. The position of a pseudo-particle is compared with the shock position. If a pseudo-particle encounters the propagating shock, the time step is reduced. And shock interaction routines described in Section 5 are performed. In this section as well as Sections 7 $\&$ 8, pseudo-particles are allowed to interact with the shock multiple times. The length of FL 2305 $s_0 \approx 2.19 R_\odot$. It takes 325 seconds for the shock to travel from the first shock-field line intersection to location VI. In Run 1, Run 2 or Run 3, we collect those pseudo-particles that cross location II, III or VI over a ten-second time interval just before the shock reaches location II, III or VI, respectively. Those pseudo-particles can be divided into accelerated pseudo-particles and escaping background pseudo-particles. Accelerated pseudo-particles are those that have interacted with the shock at least once. Escaping background pseudo-particles are those that have never interacted with the shock. The ten-second interval is considered just for getting better statistics. We compute the spectra of accelerated electrons and the spectra of escaping background electrons in a similar manner to Section 5. They are shown as red curves and blue squares in Figure \ref{fig:noreturn} respectively. We end the simulation in Step 1 when the shock hits location II, III or VI in Run 1, Run 2 or Run 3 respectively.\\

Step 2: In Run 1, Run 2 or Run 3, the shock has stopped at location II, III or VI respectively. The simulation is performed between the shock front and the visible disk at FP2 in Figure \ref{fig:loc}. All the accelerated and escaping background pseudo-particles collected in Step 1 are released at their own final positions in Step 1. They then propagate from the shock toward the solar surface in the absence of shock interaction. We solve Eqs. (26) and (27) and compute the spectra of precipitating electrons in a similar manner to Section 5. The spectra of precipitating electrons are plotted as green squares in Figure \ref{fig:noreturn}.\\

    \begin{figure}[hbtp!]
    \centering
    \includegraphics[width=1.0\linewidth]{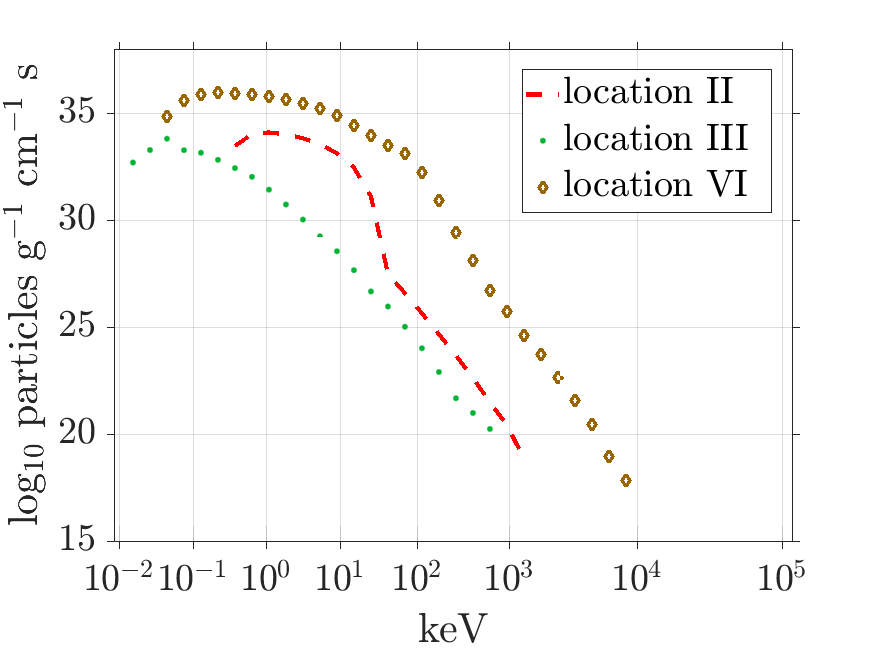}
    \caption{Same as Figure {\ref{fig:singlepreci}}, but for multiple interactions with the shock. We assume the shock only stops at locations II, III or VI rather than all the six locations (I-VI). We show the spectra of precipitating electrons, calculated from multiple interactions with the shock without particle return from downstream.
    \label{fig:noreturnpreci}}
    \end{figure}

Comparing the escaping background electron spectrum (blue squares) with the accelerated spectrum (red curve) at the three shock locations we see that significant acceleration occurs up to 43 keV by the time the shock reaches location II and hardly any particles are accelerated by the time the shock reaches location III. In contrast a huge boost in acceleration has occurred by the time the shock reaches location VI near FP2 in Figure {\ref{fig:loc}}.\\

Comparing Figures \ref{fig:noreturn} and \ref{fig:single}, we see that the inclusion in our simulations of multiple interactions produces a wider enhancement in the spectrum (up to 43 keV) than for the case for single interaction near shock location II. This is because electrons are efficiently accelerated by a quasi-perpendicular shock ($\theta_{\rm Bn}\sim84^\circ$) near this shock location. In contrast near shock location III, $\theta_{\rm Bn}\sim58^\circ$  (see Table \ref{tab:prop}) is significantly smaller than that at shock location II and the electrons do not gain much energy per SDA cycle. The strongest increase in the hardening of the electron spectrum is observed near shock location VI, shown in Figure {\ref{fig:noreturn}}~c where energies exceeding 10 MeV are produced. As the shock propagates toward FP2, many more electrons were both injected and mirrored multiple times. As the shock propagates toward location VI it also becomes increasingly quasi-perpendicular. Thus, the projected speed of the shock increases as a function of time and the shock catches the electrons it had accelerated a moment ago, even without the mirror force \citep{2006A&A...455..685S}. Particles could gain energy throughout the propagation of the shock. In order to get better statistics, we consider a ten-second time interval before the shock reaches a given location to calculate the spectra.\\

Figure {\ref{fig:noreturnpreci}} presents a comparison of the number of precipitating electrons corresponding to shock locations II, III and VI taken from Figure \ref{fig:noreturn}. As in the case for single interaction (Figure \ref{fig:singlepreci}), shock location III is providing the least precipitating electrons. Shock location VI is providing the most precipitating electrons because for each interaction with the shock the pitch angle of electrons decreases. Particles enter the loss cone and precipitate at FP2 in Figure {\ref{fig:loc}}.\\

The trapping of electrons between the moving shock and FP2 forces significant particle energy gain for as long as the particles are trapped between the two magnetic barriers i.e. until they enter the loss cone of either FP2 or the shock itself. HT resonance is also efficient in our magnetic field configurations.\\ 

\section{Particle return from downstream} \label{sec:pareturn}

Electrons transmitted downstream of the shock can scatter and return to the shock front. This is because the region behind the shock is highly turbulent and particles can easily bounce back to the upstream side of the shock. To simulate this return process from the downstream region, we follow the recipe presented in \citet{2013A&A...558A.110B}. Allowing for return from the downstream region of the shock did not change the simulation results significantly and for clarity purposes and completeness we present our approach to treat the return from downstream and the associated results in Appendix B.\\

    \begin{figure*}[hbtp!]
    \centering
    \includegraphics[width=0.75\linewidth]{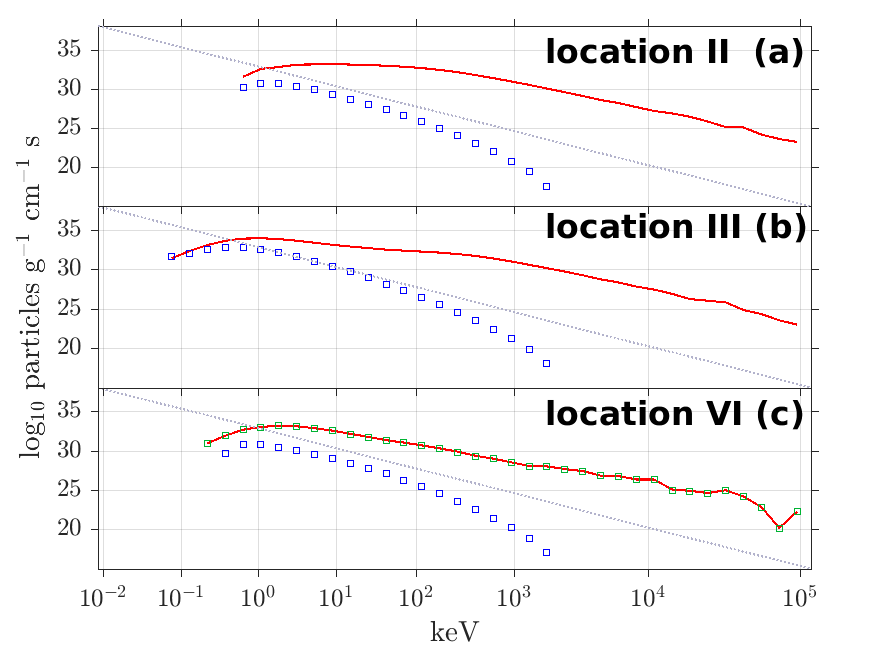}
    \caption{In the same format as Figure {\ref{fig:noreturn}}, we show the spectra of accelerated (red curves) and escaping background (blue squares) electrons calculated from DSA. For shock location VI (c), we also show the spectra of precipitating electrons (green squares). The spectra are computed for a particle scattering frequency $\nu$ = 50 Hz. The dotted gray line in each panel is a power-law with spectral index of 5 in momentum space.
    \label{fig:DSA}}
    \end{figure*}
    
\section{Diffusive-shock acceleration} \label{sec:trandiffu}

Magnetic fluctuations such as plasma waves and turbulence may exist all along coronal loops. These fluctuations may cause some level of particle scattering that will also force some particles to diffuse back to the shock for further acceleration, thus results in the so-called DSA. In addition, particle scattering can also force some particles to precipitate at FP2. These effects of diffusion are investigated in this section.\\

The level of plasma turbulence in the corona is unknown at both the large fluid scales and the small electron scales. Insights have been gained by studying the level of electron scattering in the vicinity of flaring active regions in \citet{2014ApJ...780..176K} by modeling radiation in these regions. Recent RHESSI observations of the flares of 2002 July 23, 2003 November 2, 2011 February 24 and 2011 September 24 have put constraints on the mean free path associated with the particle scattering. However, no quantitative conclusions about the strength of pitch-angle scattering have been made for the 2014 September 1 event. The level of turbulence in the quieter solar corona is even less well understood. The frequency of pitch-angle scattering $\nu$ is therefore unknown.\\ 

In this section, in Run 1 and Run 2, we only perform Step 1. In Run 3, we perform both Step 1 and Step 2. Here Step 1 is implemented in a similar manner to Step 1 in Section 7, except we now perform an isotropic small-angle scattering after we obtain the particle position $s_{\rm j}$ and pitch-angle cosine $\mu_{\rm j}$ by solving Eqs. (26) and (27) at time $t_{\rm j}$ with a fourth-order Runge-Kutta scheme. To model the DSA process, pseudo-particles are scattered through isotropic small-angle scattering at frequency $\nu$, similar to the method used in \citet{2008ApJ...675.1601A}. For each particle time step $\Delta t_{\rm j}$, the angle $\theta_{\rm s}$ between the scattering axis and the direction of propagation of the scattered particle and the azimuthal angle $\phi_{\rm s}$ around the scattering axis are randomized according to \citet{1998SoPh..182..195K}, where $\theta_{\rm s}=\sqrt{-2\nu\Delta t_{\rm j} \log{(1-\mathcal{S})}}$ and $\phi_{\rm s}=2\pi \mathcal{S'}$, and $\mathcal{S}$ and $\mathcal{S'}$ are uniformly distributed random numbers in the interval [0,1). The solution to Eq. (27) at time $t_{\rm j}$ is $\mu_{\rm j}$. Thus, the new pitch-angle cosine $\mu'_{\rm j}$ after scattering is 
\begin{eqnarray}
\mu'_{\rm j}=\mu_{\rm j}\cos\theta_{\rm s}+\sqrt{1-\mu^2_{\rm j}}\sin\theta_{\rm s}\cos\phi_{\rm s},
\end{eqnarray}
given by \citet{1990ApJ...360..702E}.\\ 

Pitch-angle scattering is performed in the entire upstream region of the shock, which can scatter electrons back to the shock. Electrons transmitted downstream of the shock can scatter and return to the shock front according to the probability of return $P_{\rm ret}$. This results in the DSA process.\\ 

In Run 3, after Step 1, we also perform Step 2 in the absence of shock interaction and scattering. Step 2 is implemented between location VI and FP2 in Run 3. The segment of FL 2305 between location VI and FP2 is very short. We ignore scattering on this very short segment in Step 2. We have to admit that this two-step process is not entirely physical but it serves as a first approximation to consider DSA.\\

\citet{2014ApJ...780..176K} found that, in coronal loops, the mean free path $\lambda\sim(10^8-10^9)$ cm for $\sim 30$ keV (e.g., the electron speed $v=10^{10}$ cm s$^{-1}$) electrons. The scattering frequency $\nu=v/\lambda$. So for $\sim 30$ keV electrons, the scattering frequency $\nu\sim10-100$ Hz. We have tried several different frequencies in the 10-100 Hz range. In Figure {\ref{fig:DSA}} we plot the spectra of accelerated and escaping background electrons for $\nu$ = 50 Hz as red curves and blue squares. We also plot the spectrum of precipitating electrons in Run 3. One can see that the inclusion of particle diffusion has given a significant boost in the acceleration of particles and precipitation rate. The spectra for all three shock locations are harder and now extend up to 100 MeV. Electrons are reflected upstream via SDA first, gaining energy necessary to be injected into DSA. Then, these reflected electrons are scattered back toward the shock for multiple cycles of acceleration. Thus, pitch-angle scattering is keeping particles in the vicinity of the shock, as expected in DSA. Pitch-angle scattering is also forcing a higher rate of precipitation by shifting particles into the mirroring loss cone of FP2 in Figure {\ref{fig:loc}} sooner.\\

Strong diffusion is more likely to occur in the downstream region than in the upstream region. This strong turbulence combined with the stronger coronal magnetic field strength compressed in the downstream region will likely lead to an increased precipitation rate at the footpoint in the shock's downstream (FP1 in Figure {\ref{fig:loc}}). We will discuss this topic in future work.\\

\section{Discussion} 
\label{sec:discuss}

The main goals of this study were to illustrate the evolution of electron acceleration at expanding shock waves and to determine whether electrons can be efficiently accelerated to relativistic energies and produce the hard X-rays measured by Fermi on 2014 September 1 shown in Figure \ref{fig:number_index} \citep{2017A&A...608A..43P}. The study employed realistic shock properties constrained by multi-point imaging \citep{2016ApJ...833...45R,2017A&A...608A..43P}, numerical models of the solar atmosphere, and a new numerical model of particle acceleration.\\

\begin{deluxetable*}{c|cc}[hbtp!]
\tablecaption{A comparison of observationally-derived and simulated numbers of $>$ 200 keV electrons and spectral indices between 11:24:32 UT and 11:24:42 UT. This corresponds to a shock passage through location VI along FL 2305. \label{tab:index}}
\tablehead{
\colhead{Acceleration process} & \colhead{Number of electrons} & \colhead{Spectral index} \\
}
\startdata
SDA (without particle return from downstream) & $6.6\times 10^{33}$ & 7.8  \\
SDA (with particle return from downstream) & $2.9\times 10^{34}$ & 7.7 \\
SDA+DSA ($\nu =$ 30 Hz) & $2.6\times 10^{34}$ & 3.2\\
SDA+DSA ($\nu =$ 50 Hz) & $1.1\times 10^{34}$  & 1.4\\
Inferred from GBM & $1.1\times 10^{32}$ & 3.6  \\
\enddata
\end{deluxetable*}

As discussed in \citet{2017A&A...608A..43P}, coronal shock waves propagate during the first hour of expansion through a complex system of magnetic loops. Figure \ref{fig:allfieldlines} presented different loops magnetically connecting the shock wave of 2014 September 1 to the visible solar disk. To illustrate the variability of shock and ambient plasma parameters along different field lines, four different loops that magnetically connected the shock to the visible disk during the hard X-ray event were considered (Figure \ref{fig:4lines}) and plotted in Figures \ref{fig:subplot1} and \ref{fig:subplot2}. These figures illustrate (1) the considerable time variability in shock properties sampled by a single magnetic loop that must be taken into account in any calculation of particle acceleration, (2) that shock properties can differ greatly along different coronal loops. In order to illustrate some important mechanisms at play, the present study focused on one of these loops only that was crossed by the shock late in the hard X-ray event but just after the peak in $\gamma$-ray emission \citep{2017A&A...608A..43P}.\\

We developed and tested a new model of particle acceleration at collisionless shocks that can take this variability into account. The model was tested first against well-established SDA theory (discussed in Appendix A). Combining the particle acceleration code with the shock modeling we determined for a single magnetic field line (FL 2305) the important processes at play during electron acceleration to high energy.\\

The combination of the loop topology and the changing location of the shock-loop connection means that the shock geometry can evolve very rapidly along a single field line. SDA was efficient when the shock first intersected the loop (see Figure {\ref{fig:single}}). As the shock reached the loop top it temporarily became more quasi-parallel where few electrons could be accelerated by a single SDA cycle. An additional factor is related to the pool of particles available for acceleration. When the shock intersects the loop in the dense plasma prevalent in the low corona, the number of electrons available for acceleration is necessarily higher than near the more tenuous loop top.\\

We proceeded by allowing for multiple encounters with the shock, which clearly increased the number of accelerated and precipitating electrons (demonstrated in Figure {\ref{fig:noreturn}}). The effect of the mirror force and the resulting precipitation of electrons (see Figure {\ref{fig:noreturnpreci}}) were similar to the previous case. We then allowed electrons transmitted through the shock to return from downstream (Appendix B). The number of accelerated electrons did not increase significantly, but slightly more electrons were able to precipitate (see Figures {\ref{fig:return}} and {\ref{fig:returnpreci}}). Finally, we moved on to the acceleration of electrons through DSA. As demonstrated in Figure {\ref{fig:DSA}}, particle scattering can boost the acceleration of electrons by forcing additional interactions with the shock and the precipitation rate by shifting particles into the loss cone of FP2 in Figure {\ref{fig:loc}}.\\

As mentioned above, particle scattering frequencies $\nu$ are not known in the corona and only estimates have been inferred from past modeling of solar flares. We allow the shock to propagate through the entire loop to location VI (which is very close to FP2) and sum accelerated and escaping background electrons during the last 10 seconds just before the shock wave reaches location VI at 11:24:42 UT. Thus, accelerated and escaping background electrons that reach shock location VI from 11:24:32 UT to 11:24:42 UT are summed. We then propagate them to the chromosphere and count the number of precipitating electrons $N_{\rm pr}$.\\

To evaluate whether the simulated number of precipitating electrons could even roughly explain the hard X-rays measured by Fermi GBM during the 2014 September 1 event, we compare the number of simulated electrons that precipitate along FL 2305 with the number of electrons inferred from the hard X-ray spectra \citep{2017A&A...608A..43P}. The hard X-rays measured by GBM are integrated over the entire visible disk. For a meaningful comparison we must therefore consider how the shock connects magnetically to the visible disk via many different realistic flux tubes. In this study, we assume that all field lines connecting the shock to the visible disk are like FL 2305. Since the shock properties along this particular loop were highly favorable for particle acceleration our estimate of the number of energetic electrons precipitating at the surface of the Sun is likely to be an upper limit since most other loops will be less efficient.\\

The area of the solar surface in the shock's upstream covered by the magnetic flux tube surrounding FL 2305 is assumed to be 1 cm$^2$. According to the bottom panel of Figure 11 in \citet{2017A&A...608A..43P}, about 2$\%$ of the total visible solar surface is magnetically connected to the shock, which is 2$\%$ $\times$ 2 $\times$ $\pi$ $\times$ $R^2_\odot \approx$ 6 $\times$ 10$^{20}$ cm$^2$. Assuming that all loops connecting to the visible disk are like FL 2305, there would be about $N_{\rm pr}$ $\times$ 6 $\times$ 10$^{20}$ precipitating electrons from 11:24:32 UT to 11:24:42 UT.\\

We fit Fermi GBM spectra using a thick target electron-proton bremsstrahlung model and obtain $1.13 \times 10^{32}$ electrons above 200 keV from 11:24:32 UT to 11:24:42 UT. We report the simulated numbers of precipitating electrons and compare them to the number of electrons inferred from GBM in Table {\ref{tab:index}}.\\

Simulations that take multiple interactions by SDA into account provide more than enough particles to explain the number of electrons inferred from GBM. For DSA simulations, shown in Figure {\ref{fig:DSA}} we assumed a particle scattering frequency of $\nu=50$ Hz. For $\nu=$ 50 Hz, we found that $1 \times 10^{34}$ electrons precipitated with energies greater than 200 keV between 11:24:32 UT to 11:24:42 UT. We ran a similar simulation assuming a lower frequency of $\nu=30$ Hz for which the number of precipitating electrons is reported in Table {\ref{tab:index}} as well. A value $\nu$ = 30 Hz is also reasonable in the loop \citep{2014ApJ...780..176K}.\\

We remind the reader that the comparison between observed and modeled precipitating electrons is derived from only one loop with very favorable conditions, and this comparison does not consider the variability induced by different loops. In a future study, the numbers may change when we consider all loops but this first simple comparison is very promising.\\

Estimates of the number of electrons from our simulations are interesting but prone to considerable uncertainty since we have not accounted yet for the variability of all field lines. A unique feature of GBM hard X-ray data is the relatively constant and hard power-law spectrum throughout the event shown in Figure \ref{fig:number_index}c. Our single loop model provides good ground for comparison with such a spectrum. The right-hand column of Table {\ref{tab:index}} presents the spectral indices derived from the different simulated spectra. The spectral indices derived from SDA simulations are greater than 7 even when multiple encounters are allowed and return from downstream is taken into account. Simulations that fold in diffusion (DSA) on the other hand are able to produce very hard spectra. In particular for a particle scattering frequency of $\nu=30$ Hz, the spectral index (3.2) is very close to the observed one (3.6).\\ 

The analysis made the most of available coronal observations. We have obtained realistic shock reconstructions at high cadence that address to some extent the variability of acceleration processes along a magnetic loop. We have used realistic ambient coronal conditions derived from established benchmarked MHD simulations. There are nevertheless a number of limitations in the approaches taken in this work. The pitch-angle scattering frequency $\nu$ is unknown and will remain for some time an open question, and as already mentioned we have performed our simulations on only one magnetic loop. A future study must extend the analysis to calculations of particle acceleration and transport along multiple magnetic field lines. We have also ignored cross-field diffusion and electron transport between adjacent loops that might be possible in response to either a dynamic reconfiguration of coronal loops or a significant level of particle scattering.\\ 

\section{Conclusions} 

Our model roughly reproduces the very hard electron spectrum and the number of electrons inferred from Fermi GBM. Thus, we have obtained more evidence for the CME-driven shock being an important accelerator of energetic electrons in the corona to produce hard X-rays far away from the flare. In this study we have not addressed acceleration on open magnetic field lines that are connected to interplanetary space and potentially to probes situated in the inner heliosphere. Some electrons may be accelerated by the same shock along open magnetic fields and produce SEPs measured in situ. For the particular event studied here, \citet{2017A&A...608A..43P} reported the occurrence of a strong SEP event with 2.8-4.0 MeV electrons measured at STEREO-B. Accelerating electrons to 2.8-4.0 MeV along open magnetic field lines will require a single open field line intersecting with the shock front multiple times \citep{2019Natur.576..237B, 2019Natur.576..228K}. We have found such cases in our shock modeling for this event. This mechanism will also be quantified in a future study. Alternatively, magnetic fluctuations may trap electrons in the vicinity of the shock, which leads to efficient DSA acceleration. In the presence of magnetic disturbances, electrons can reach 2.8-4.0 MeV along open magnetic field lines. Or potentially, electrons are accelerated along closed magnetic loops first. Then these closed loops reconnect to open field lines, which allows those accelerated energetic electrons to escape to the interplanetary space. Another possibility is that electrons first gain energy along closed loops, then reach open field lines via cross-field diffusion jump. This will be studied further in a future study.\\

In coming years, Parker Solar Probe and Solar Orbiter will no doubt measure many energetic particle events closer to the Sun in times of elevated electromagnetic radiation (which also will be measured with the STIX instrument on Solar Orbiter). This study is an excellent preparation for future studies that will compare the populations of energetic particles released from the solar corona with those that impact the solar surface.\\

\acknowledgments

We acknowledge Gerald H. Share (Department of Astronomy, University of Maryland, College Park, MD 20742, USA) for providing the number and the power-law spectral index of electrons inferred from the Fermi/GBM observations. We also thank the referee for several helpful comments. The IRAP team acknowledges support from the French space agency (Centre National des Etudes Spatiales; CNES; \url{https://cnes.fr/fr}). Y.W. and A.K. acknowledge financial support from the ANR project SLOW{\_}\,SOURCE (ANR-18-ERC1-0006-01), COROSHOCK (ANR-17-CE31-0006-01), and FP7 HELCATS project \url{https://www.helcats-fp7.eu/} under the FP7 EU contract number 606692. The work of A.P.R. was funded by the ERC SLOW{\_}\,SOURCE project (SLOW{\_}\,SOURCE - DLV-819189). The work at the University of Turku was conducted in the framework of the Finnish Centre of Excellence in Research of Sustainable Space, funded by the Academy of Finland (grant 312357). R.V. and A.A. also acknowledge the financial support of the Academy of Finland of the project 309939. The work of A.W. was supported by DLR grant No. 50 QL 1701. Support by ISSI and ISSI-BJ through the international teams 469 is also acknowledged.

%


\appendix 

\section{Test of the Monte Carlo simulation}

\begin{figure*}[hbtp!]
    \epsscale{1.3}
    \plotone{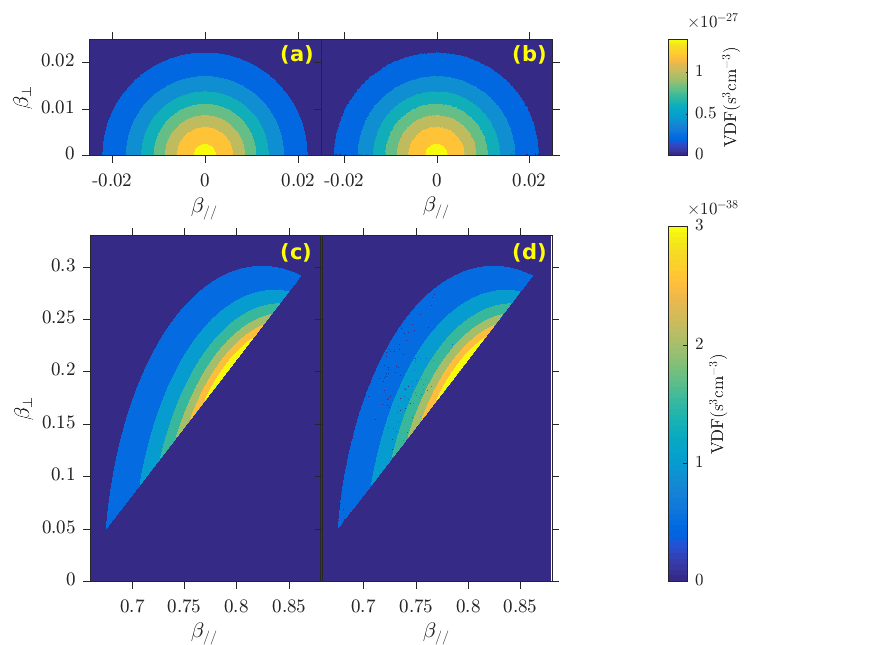}
    \caption{Electron initial VDFs: analytical $f_0$ (Panel a) and simulated initial VDF (Panel b). Electron VDFs after one reflection: analytical $f_{\rm acc}$ (Panel c) and simulated VDF (Panel d).
    \label{fig:f2d}}
    \end{figure*}

In this section we assume the initial distribution function is an isotropic $\kappa$-distribution function of momentum $p$ given by
\begin{eqnarray}
f_{0}(p)=\frac{1}{(\pi\kappa p^2_\kappa)^{3/2}}\cdot\frac{\Gamma(\kappa+1)}{\Gamma(\kappa-1/2)} \cdot\left[1+\frac{p^2}{\kappa p^2_\kappa}\right]^{-\kappa-1}
\end{eqnarray}
where $p_\kappa={m_0 w_\kappa}/{\sqrt{1-{w^2_\kappa}/{c^2}}}$ is the momentum corresponding to the thermal velocity $w_\kappa$ (cf Eq. (17) in \citet{2009A&A...494..669M}). And $w_\kappa=\sqrt{(2\kappa-3)k_{\rm B} T/\kappa m_0}$ with $k_{\rm B}$ being the Boltzmann constant, $T$ being the temperature and $m_0$ being the rest mass of the particle. Eq. (A1) can be transformed to a velocity distribution function (VDF) as
\begin{eqnarray}
f_0(\beta_{\rm i,\parallel},\beta_{\rm i,\perp})=\left(\frac{\pi\kappa w^2_\kappa}{c^2-w^2_\kappa}\right)^{-3/2}\cdot\frac{\Gamma(\kappa+1)}{\Gamma(\kappa-1/2)}\nonumber \\
\times\left\{1+\frac{(c^2-w^2_\kappa)(\beta^2_{\rm i,\parallel}+\beta^2_{\rm i,\perp})}{\kappa w^2_\kappa[1-(\beta^2_{\rm i,\parallel}+\beta^2_{\rm i,\perp})]}\right\}^{-\kappa-1}.
\end{eqnarray}
The electron speed is randomized from the $\kappa$-distribution. In Figure \ref{fig:f2d}~b we present the simulated initial electron VDF with parameters: temperature $T$ = 1.5 MK and $\kappa$ = 3. We also assume $\theta_{\rm Bn}$ = 89.7$^\circ$, $V_{\rm sn}$ = 1000 km/s and $\gamma_{\rm B}$ = 2. If an initial electron satisfies the reflection conditions Eqs. (21) and (22), we reflect this electron. In Figure \ref{fig:f2d}~d we show the simulated VDF of these electrons that experience a single reflection.\\

In order to validate our Monte Carlo code, we solve analytically the VDF of the electrons that experience a single reflection. We consider the same initial electron VDF Eq. (A2) with the same parameters: temperature $T$ = 1.5 MK and $\kappa$ = 3. In Figure \ref{fig:f2d}~a we present the analytical initial electron VDF. From Eqs. (21) and (22), we can derive the analytical VDF of the accelerated electron is a shifted loss-cone distribution \citep{2006A&A...454..969M, 2009A&A...494..669M}
\begin{eqnarray}
f_{\rm acc}=\Theta\left(\beta_{\rm s}-\beta_{\rm i,\parallel}\right)\nonumber \\
\times\Theta\left(\beta_{\rm i,\perp}-\frac{\tan \alpha_{\rm c}}{\sqrt{1-\beta^2_{\rm s}}}\cdot(\beta_{\rm s}-\beta_{\rm i,\parallel})\right)\nonumber \\
\times f_0(\beta_{\rm i,\parallel},\beta_{\rm i,\perp}),
\end{eqnarray}
where $\Theta$ is the Heaviside step function and $f_0$ denotes the initial VDF (cf Eq. (5) in \citet{2009A&A...494..669M}). Thus, the resulting analytical VDF of the electrons that experience a single reflection is
\begin{eqnarray}
f_{\rm acc}=\left(\frac{\pi\kappa w^2_\kappa c^2}{c^2-w^2_\kappa}\right)^{-3/2}\cdot\frac{\Gamma(\kappa+1)}{\Gamma(\kappa-1/2)}\nonumber \\
\times\Theta\left(\beta_{\rm s}-\frac{2\beta_{\rm s}-\beta_{\rm r,\parallel}(1+\beta^2_{\rm s})}{1-2\beta_{\rm r,\parallel}\beta_{\rm s}+\beta^2_{\rm s}}\right)\nonumber \\
\times\Theta\left(\frac{(1-\beta^2_{\rm s})\beta_{\rm r,\perp}}{1-2\beta_{\rm r,\parallel}\beta_{\rm s}+\beta^2_{\rm s}}\right.\nonumber \\
\left.-\frac{\tan \alpha_{\rm c}}{\sqrt{1-\beta^2_{\rm s}}}\cdot\left[\beta_{\rm s}-\frac{2\beta_{\rm s}-\beta_{\rm r,\parallel}(1+\beta^2_{\rm s})}{1-2\beta_{\rm r,\parallel}\beta_{\rm s}+\beta^2_{\rm s}}\right]\right)\nonumber \\
\times\left\{1+\frac{c^2-w^2_\kappa}{\kappa w^2_\kappa(1-\beta^2_{\rm s})^2(1-\beta^2_{\rm r,\perp}-\beta^2_{\rm r,\parallel})}\right.\nonumber \\
\times\left[(1-2\beta_{\rm r,\parallel}\beta_{\rm s}+\beta^2_{\rm s})^2\right.\nonumber \\
\left.\left.-(1-\beta^2_{\rm s})^2(1-\beta^2_{\rm r,\perp}-\beta^2_{\rm r,\parallel})\right]\right\}^{-\kappa-1}.
\end{eqnarray}
In Figure \ref{fig:f2d}~c we show this analytic VDF of the electrons after one reflection calculated with the same parameters $\theta_{\rm Bn}$ = 89.7$^\circ$, $V_{\rm sn}$ = 1000 km/s and $\gamma_{\rm B}$ = 2. We compare the analytical and simulated VDFs of the initial electrons and the electrons that experience a single reflection in Figure \ref{fig:f2d}. The VDFs in the left column and the right column are identical, as expected.\\

\section{Allowing for return from the downstream region:}

The downstream plasma is assumed to be very turbulent so that the fate of the particle is basically determined instantly and based on local fluid properties. For isotropic downstream populations, the probability of return to the shock front was given in \citet{1991SSRv...58..259J} as
\begin{eqnarray}
P_{\rm ret}  & = &
 \left\{
\begin{array}{ll}
\left(\frac{v'-u_2}{v'+u_2}\right)^2, & v'>u_2 \\
0, & v'\le u_2,
\end{array}     \right. 
\end{eqnarray}
where $v'$ is the particle velocity in the downstream rest frame and $u_2$ is the plasma speed in the downstream region in the HT frame. Successfully returning electrons re-enter the upstream region, and may experience energy gains.\\

A similar simulation as presented in Section 6 is repeated here, except we now modify shock interaction routines to allow pseudo-particles transmitted through the shock to return from downstream. If a pseudo-particle fulfills Eq. (21), but it is inside the loss cone of the shock, we calculate the return probability $P_{\rm ret}$. If $P_{\rm ret}>0$, we return this pseudo-particle back to the upstream side of the shock and adjust its weight by multiplying its original weight with $P_{\rm ret}$.\\

    \begin{figure*}[hbtp!]
    \centering
    \includegraphics[width=0.75\linewidth]{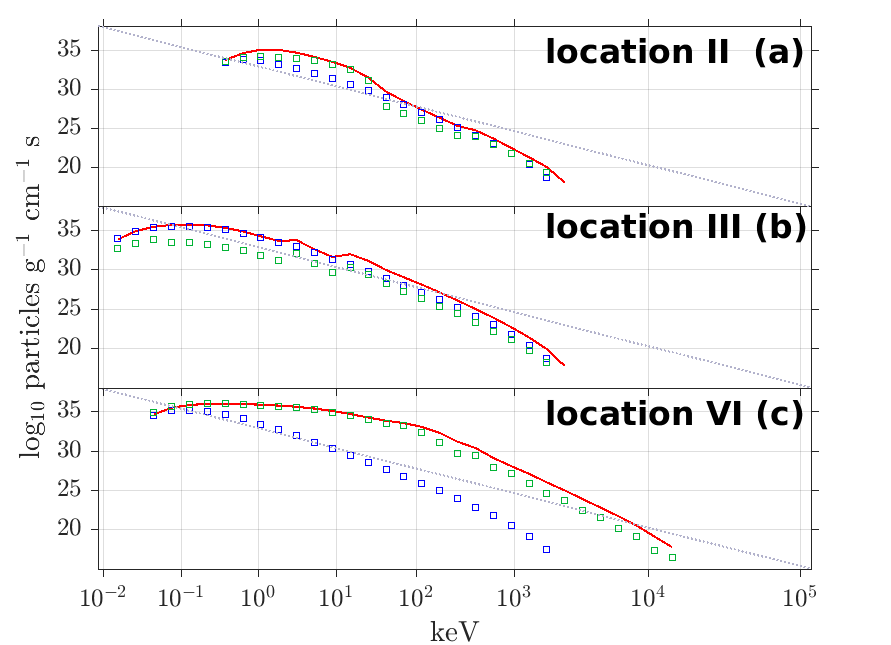}
    \caption{Same as Figure {\ref{fig:noreturn}}, but calculated from multiple interactions with the shock with particle return from downstream. The dotted gray line in each panel is a power-law with spectral index of 5 in momentum space.
    \label{fig:return}}
    \end{figure*}
    
    \begin{figure}[hbtp!]
    \centering
    \includegraphics[width=1.0\linewidth]{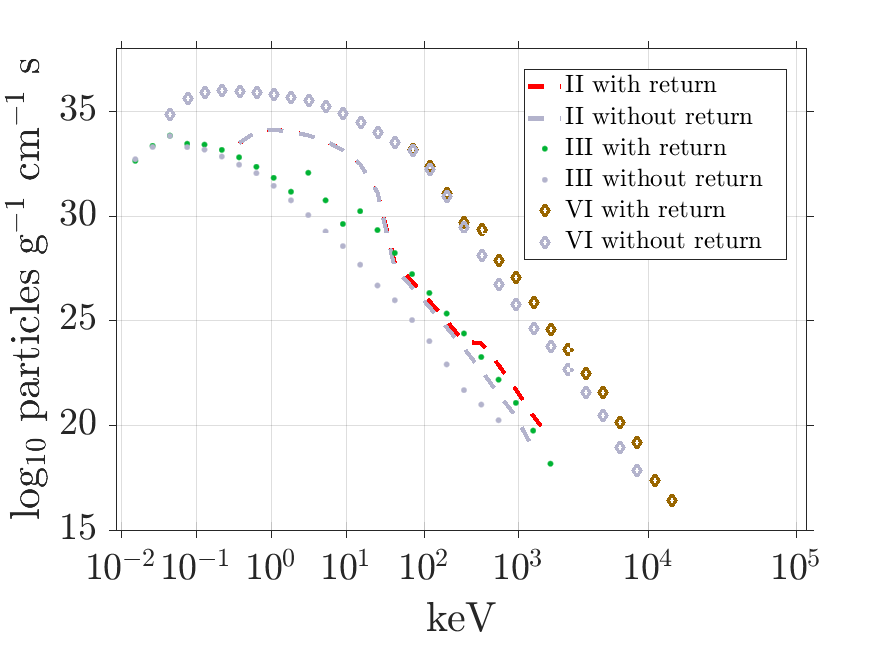}
    \caption{In the same format as Figure {\ref{fig:noreturnpreci}}, we show the spectra of precipitating electrons, calculated from multiple interactions with the shock with particle return from downstream, in dashed red curve, green dots and brown diamonds. For comparison, we show the spectra of precipitating electrons for multiple interactions with the shock without particle return from downstream in gray.
    \label{fig:returnpreci}}
    \end{figure}

In Figure {\ref{fig:return}}, we plot the spectra of accelerated, escaping background and precipitating electrons with the same scheme as before. It should be noted that returning from downstream is now taken into account.\\

When particles are allowed to return from downstream, slightly more particles precipitate from all shock locations compared with simulations when particles are not allowed to return (see Figure {\ref{fig:returnpreci}}). This is due to high-energy particles that, in the previous set of simulations were crossing the shock and removed from the simulation. Return from downstream being a diffusive process, these high-energy particles are sent back to the upstream region with the appropriate modifications to speed and pitch angle. If their new pitch angles are small enough, they can enter the loss cone and precipitate at FP2 in Figure {\ref{fig:loc}}. Figure {\ref{fig:returnpreci}} shows that the particles released from all three shock locations can now precipitate at high energies. These high-energy particles were in the previous simulations but were lost behind the downstream region and are now returned to the shock.\\

\end{document}